\documentclass[10pt,twocolumn]{IEEEtran}
\usepackage{times,amsmath,epsfig}
\usepackage{xspace,latexsym,syntonly}
\usepackage{amssymb}
\usepackage{amsfonts}
\usepackage{textcomp}
\usepackage{graphicx}
\usepackage{epsfig}

\newcommand{\SNR}{\ensuremath{\hbox{SNR}}}

\newcounter{theorem}
\newtheorem{theorem}{Theorem}
\newtheorem{lemma}[theorem]{Lemma}
\newtheorem{claim}[theorem]{Claim}
\newcounter{definition}
\newtheorem{definition}{Definition}

\newcommand{\beq}{\begin{equation}}
\newcommand{\eeq}{\end{equation}}
\newcommand{\bea}{\begin{array}}
\newcommand{\ena}{\end{array}}
\newcommand{\bds}{\begin {itemize}}
\newcommand{\eds}{\end {itemize}}
\newcommand{\bdf}{\begin{definition}}
\newcommand{\blm}{\begin{lemma}}
\newcommand{\edf}{\end{definition}}
\newcommand{\elm}{\end{lemma}}
\newcommand{\bthm}{\begin{theorem}}
\newcommand{\ethm}{\end{theorem}}
\newcommand{\bprp}{\begin{prop}}
\newcommand{\eprp}{\end{prop}}
\newcommand{\bcl}{\begin{claim}}
\newcommand{\ecl}{\end{claim}}
\newcommand{\bcr}{\begin{coro}}
\newcommand{\ecr}{\end{coro}}
\newcommand{\bquest}{\begin{question}}
\newcommand{\equest}{\end{question}}

\newcommand{\larrow}{{\larrow}}

\begin{document}
\title{Fully distributed optimal channel assignment for open spectrum access}
\author{Oshri Naparstek ,Student Member, IEEE and Amir Leshem ,Senior Member, IEEE\thanks{Copyright (c) 2013 IEEE. Personal use of this material is permitted. However, permission to use this material for any other purposes must be obtained from the IEEE by sending a request to pubs-permissions@ieee.org.
O. Naparstek is with the School of Engineering, Bar-Ilan University, Ramat-Gan,
52900, Israel (e-mail: oshri8@gmail.com).
A. Leshem is with the School of Engineering, Bar-Ilan University,
Ramat-Gan, 52900, Israel.
This research was partially supported by the CORNET consortium of the Israeli MAGNET.
Parts of this paper had been presented in \cite{naparstek2011},\cite{naparstek2012}}}
%\affiliation{School of Engineering, Bar-Ilan University, Ramat-Gan, Israel}
\date{\today}
\maketitle
\begin{abstract}
In this paper we address the problem of fully distributed assignment of users to sub-bands such that the sum-rate of the system
 is maximized. We introduce a modified auction algorithm that can be applied in a fully distributed way using an opportunistic CSMA assignment scheme and is $\epsilon$ optimal.
  We analyze the expected time complexity of the algorithm and suggest a variant to the algorithm that has lower expected complexity.
We then show that in the case of i.i.d Rayleigh channels a simple greedy scheme is asymptotically optimal as $\SNR$ increases or
 as the number of users is increased to infinity.
  We conclude by providing simulated results of the suggested algorithms.
\end{abstract}
\section{Introduction}
\label{section_introduction}
Recently, there had been an accelerated deregulation of spectrum usage. This can primarily be attributed to the success of  communication in the unlicensed $2.4$GHz band and the large markets created by devices operating there.
A key factor in the success of the mass deployment of devices in the unlicensed bands is their ability to sense the spectrum and transmit over various frequency bands.
The technology that enables the coexistence of different devices over the same frequency band is called cognitive radio.
There are three main models for spectrum sharing in cognitive radio systems. The hierarchial model, open sharing model and dynamic exclusive use model. The hierarchial model, where secondary users are allowed to use the spectrum when the primary users are not is the best known. In this paper we focus on the open spectrum access model where there are no primary users in the network. A good overview of the various models of dynamic spectrum access is given by Zhao and Sadler \cite{zhao2007survey}.
Here, cognitive radio systems  are defined as radio systems operating over multiple frequency selective wireless channels in which users can change their transmission or reception parameters to communicate without interfering with other users.
One way to avoid interference between users sharing the same frequency band is to split the spectrum into $K$ orthogonal
sub-bands using Orthogonal Frequency Division Multiple Access (OFDMA). If users can be assigned to sub-bands efficiently gains can be derived from the diversity of the channel. This problem is known as the channel assignment problem. The centralized assignment of sub-carriers to users has been addressed extensively over the last decade because of the high demand for efficient spectrum utilization in wireless and wireline communication systems. Papers that consider joint power and sub-carrier allocation in the downlink direction include \cite{yui1999}-\nocite{kim2006}\cite{Shen2005}. Sub-carrier allocation methods in the uplink direction include \cite{kim2005}-\nocite{gao2008}\nocite{tang2009}\cite{huang2009}.

The channel assignment problem is a special case of the assignment problem. The assignment problem is a classical optimization problem. The original formulation of the assignment problem is as follows: Given a matrix $\textbf{A}$ find a permutation matrix $\textbf{P}$ that maximizes the trace $tr(\textbf{AP})$. The first specialized algorithm to solve the assignment problem was the Hungarian method suggested by Kuhn in $1955$ \cite{kuhn1955hungarian}. Later, this method was generalized by Dantzig to solve general linear programming problems \cite{dantzig1956primal}. All of the methods to solve the assignment problem were centralized and required full knowledge of the utility matrix. In 1979 a distributed relaxation of the assignment problem called the auction algorithm was introduced by Bertsekas \cite{bertsekas1979distributed}. It is called the auction algorithm since it was inspired by auctions where the users bid for objects and raise their bids until the highest bidder wins the object. The auction algorithm did not require full knowledge of the utility matrix but did need some kind of explicit message passing mechanism or a shared memory. Furthermore, the auction algorithm was shown to converge to a solution which is at most $N\epsilon$ from the optimal solution where $N$ is the number of users and $\epsilon>0$ is a small constant chosen by the user. Furthermore, when utilities are integers and $\epsilon < \frac{1}{N}$ the assignment is optimal.
The auction algorithm has previously been suggested as a way to solve the channel assignment problem.
In \cite{yang2009} the auction algorithm was used to solve the channel assignment problem for wireless networks in the uplink direction. In \cite{bayati2007} a distributed auction algorithm with shared memory was used as a solution for switch scheduling. In \cite{bayati2008} it was shown that a modification of the auction algorithm is equivalent to the max product belief propagation. However, these modified auction algorithms also required shared memory. Note that there is another class of algorithms that are also called auction algorithms. These algorithms rely mainly on economic principles and game theory and are mainly aimed at fairness problems. Papers that consider this type of auction algorithms include \cite{jun2006}-\nocite{gai2009}\nocite{han2011}\nocite{kwon2010}\cite{hung2010}. This type of auction algorithm is not considered in this paper.

Another field of research dealing with optimal assignments is the field of random assignments. Unlike the deterministic approach to the assignment problem that is mainly concerned with developing methods to find an optimal assignment, in random assignment problems the statistical properties of the optimal assignments are studied. The most widely known result in this area is the
conjecture by Me�zard and Parisi \cite{mezard1985replicas} that the expected value of a random linear minimum assignment problem with i.i.d uniform $[0, 1]$ or exponential with a mean $1$ cost coefficients approaches $\frac{\pi^2}{6}$ as the size of the problem goes to infinity. The conjecture was proven by Aldous \cite{aldous2001zeta} in $2001$. A simpler proof was found by Linusson and W{\"a}stlund \cite{linusson2004proof} in $2004$. A good review of the results on random assignment problems is given in \cite{krokhmal2009random}. We use Aldous's theorem to provide tight estimates on the expected time complexity of the auction algorithm.

Unlike cellular and optical systems that have centralized access management, cognitive radio systems are
inherently distributed. In this case, centralized optimization methods cannot be used and distributed channel assignment protocols are needed to assign the users to the sub-channels. The simplest protocol to implement in the distributed case is a random channel assignment. However, it was shown in \cite{leshem2011a} that for a large number of users, the relative loss of the random allocation is at most $\frac{1}{\log K}$, where $K$ is the number of channels. Recently there has been growing interest in spectrum optimization for
frequency selective channels. However, most of the work done in this field relies on explicit exchange of channel state information or channel statistics. Several suboptimal approaches that do not require information sharing have been suggested for the channel assignment problem. In \cite{kwon20102} a greedy approach to the channel assignment problem was introduced.
In \cite{leshem2011a},\cite{yaffe2010} the use of opportunistic carrier sensing was combined with the Gale Shapley algorithm for stable matching \cite{gale1962college} to provide a fully distributed stable channel assignment. That solution basically achieves a greedy channel assignment.

In this paper we generalize the solution in \cite{leshem2011a} to the more general case of fully distributed maximum weighted sum-rate channel assignment where no explicit message passing or shared memory is possible. We suggest a fully distributed scheme that relies on a modified auction algorithm to solve the distributed channel assignment problem using an opportunistic multi channel Carrier Sense Multiple Access (CSMA) protocol. The suggested scheme does not require any explicit message passing mechanism or shared memory and converges to a solution within $N\epsilon$ from the optimal solution where $N$ is the number of users and $\epsilon$ is a small scalar. We analyze the expected complexity of the distributed auction algorithm. An efficient variant of the distributed auction algorithm termed the truncated distributed auction algorithm is introduced. We show that the truncated auction algorithm is asymptotically optimal when the number of users is large enough. We then analyze the random channel assignment problem under Rayleigh channels. We show that a randomized greedy assignment that converges within one time frame is asymptotically optimal in both high $\SNR$, and as the number of users increases.

The  paper is organized as follows: In section \ref{section_problem_formulation} we discuss our model for the cognitive radio system and the spectrum allocation problem. In section \ref{section_auction_alg} we provide a brief overview of the auction algorithm and introduce the fully distributed auction algorithm that relies solely on local information. We show that the modified auction algorithm shares some important properties with the original auction algorithm such as the bound on optimality. In section \ref{section_auction_csma} we show how the modified auction algorithm and opportunistic CSMA can be combined to solve the distributed channel assignment problem without using shared memory or a control channel. In section \ref{section_ub_iter} we analyze the expected time complexity of the distributed auction algorithm. In section \ref{section_performance} we suggest an asymptotically optimal variant to the distributed auction algorithm that has a lower expected time complexity. In section \ref{section_opt_rayleigh} we analyze a randomized greedy assignment scheme for i.i.d Rayleigh fading channels and show that it is asymptotically optimal. Finally we present some simulation results and conclusions.

\section{Problem formulation}
\label{section_problem_formulation}
\begin{figure}[htbp]
\centering \includegraphics[width=0.5\textwidth]{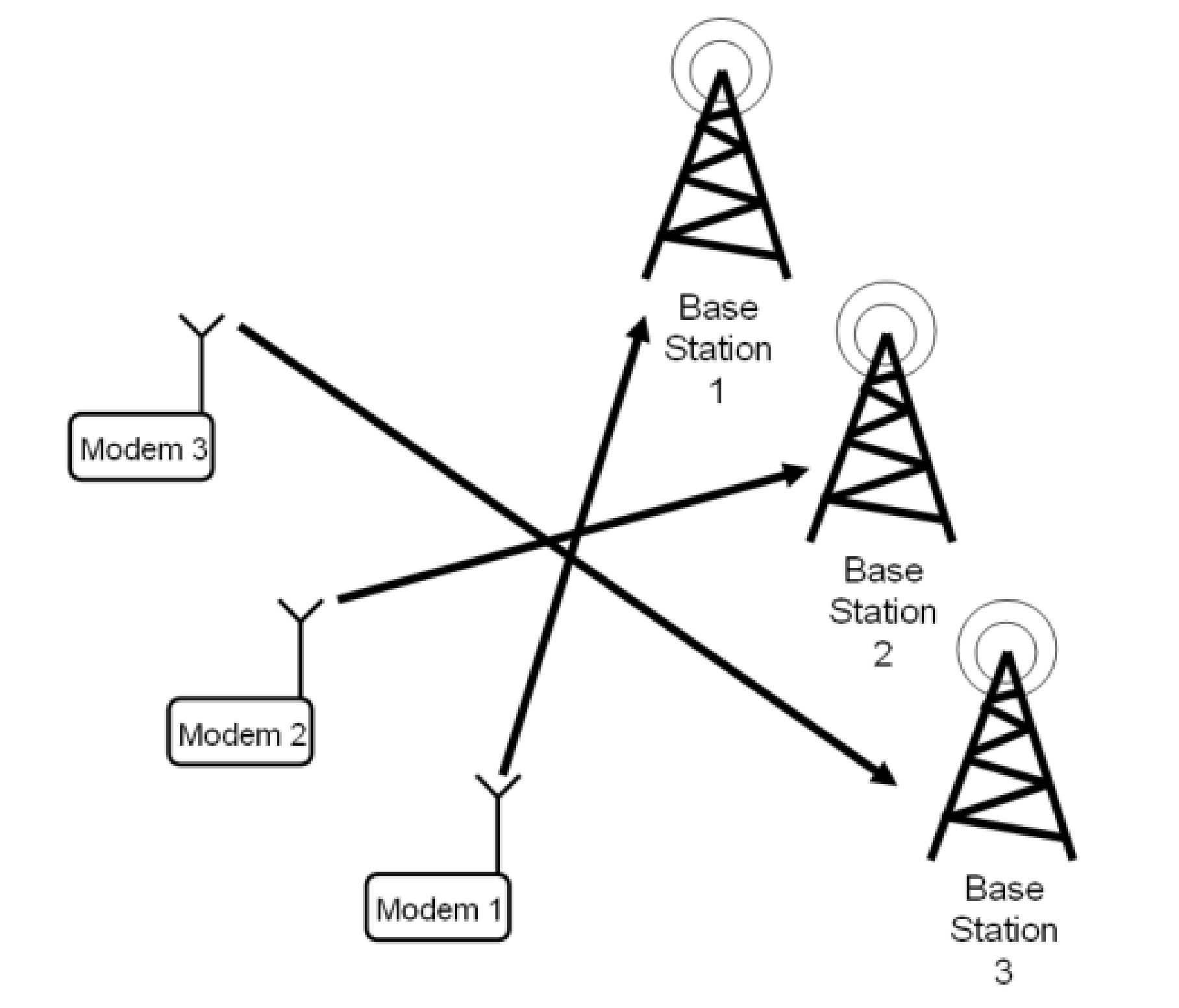}
\caption{Cognitive radio pairs using an unlicensed band}
\label{figure_setup}
\end{figure}

We consider a time slotted system of $N$ cognitive users using an unlicensed spectrum band divided into $K$ sub-channels.
%This assumption is always true because if $M>K$ than $K-M$ artificial channels with rate $0$ could be added and make $M=K$ \cite{bertsekas1991reverse}.
We assume all the users have continuous sensing over all channels.
$\textbf{R}(n,k)$ is defined to be the average weighted rate of user $n$ in channel $k$.
\beq
\textbf{R}(n,k)=w_n\log_2(1+\SNR(n,k))
\eeq
where $w_n$ is the weight of user $n$ and $\SNR(n,k)$ is the signal to noise ratio of the $n$'th user in the $k$'th channel.
We assume that only one user can transmit on each channel at each time slot and consider out of cell interferences as noise.
An illustration of the system model is given in Fig. \ref{figure_setup}.
We are aiming for a distributed algorithm that finds an assignment that maximizes the sum-rate of all users.
The maximum weighted sum-rate  problem can be formulated as an integer programming problem.
\beq\label{Assign_LP12}
\bea{ll}
\displaystyle\max \sum_m\sum_k \textbf{R}(n,k)\boldsymbol\eta(n,k)\\
 s.t \\
 \sum_k\boldsymbol\eta(n,k)=1,&\forall n  \\
  \sum_n\boldsymbol\eta(n,k)=1,&\forall k  \\
  \boldsymbol\eta(n,k)\in \{0,1\} ,&\forall n,k
\ena
\eeq
where $\eta(n,k)\in \{0,1\}$ equals 1 if the $k$'th channel is assigned to the $n$'th user.
The constraint matrix of this problem is totally unimodular. This means that the solution to the relaxed problem where we replace the integer constraint by $\eta(n,k)\geq 0$ is also the solution to the original problem. The relaxed problem is a linear programming (LP) problem and can be solved efficiently by LP solutions methods. The assignment problem can be solved by specialized methods such as the Hungarian Method \cite{kuhn1955hungarian}. However, most of the methods that solve the assignment problem cannot be used easily in a parallel and distributed manner.

\section{The auction algorithm}
\label{section_auction_alg}
The auction algorithm \cite{bertsekas1979distributed} is an intuitive method for solving the assignment problem. In many cases it has been shown to converge faster than other methods for solving the assignment problem \cite{bertsekas1990auction}. Another advantage over other methods is that it can be used in a distributed or parallel manner very easily. Auctions in which unassigned people raise their prices and bid for objects simultaneously was the original inspiration for the auction algorithm. Similarly, the auction algorithm has two stages, the bidding stage and the assignment stage. In the bidding stage each unassigned individual raises the price of the object he wishes to acquire by the difference between the most profitable object and the second most profitable object plus some constant $\epsilon$. In the assignment stage every object is assigned to the highest bidder. The two stages are repeated until all bidders are assigned an object. More specifically, let $\textbf{R}$ be the matrix of the initial rewards. Let $\textbf{B}$ be a matrix of the bids. $\boldsymbol\rho$ is the price vector where $\rho_k$ is given by:
\beq\label{eq_rho}
\rho_k=\max_{n}\textbf{B}(n,k)
\eeq
Let $\boldsymbol\eta=\left[\eta_1,\eta_2,...,\eta_N\right]$ be an assignment (permutation) vector  where $\eta_n$ is the object assigned to person $n$.
\definition An object $k$ is said to be \emph{assigned} to person $n$ by $\boldsymbol\eta$ if $\boldsymbol\eta_n=k$.
The reward of the $n$'th person on assignment $\boldsymbol\eta$ is denoted by $\textbf{R}(n,\eta_n)$
and the price that the $n$'th person pays on assignment $\boldsymbol\eta$ is denoted by
$\rho_{\boldsymbol\eta_{n}}$.
Given a positive scalar $\epsilon$, an assignment $\eta$ and a price $\rho_{\eta_n}$, a person $n$ is termed \emph{happy} with assignment $\boldsymbol\eta$ if the profit (i.e., reward minus price) is within $\epsilon$ of the maximal profit achievable by person $n$. This condition is called $\epsilon$-Complementary slackness ($\epsilon$-CS).
\beq\label{eq_ecs}
\textbf{R}(n,\eta_n)-\rho_{\eta_n}\geq\max_k(\textbf{R}(n,k)-\rho_k)-\epsilon
\eeq
When all the people are assigned and happy, the algorithm stops. It was shown in \cite{bertsekas1979distributed} that the algorithm terminates in finite time. The algorithm is within $N\epsilon$ of being optimal at termination \cite{bertsekas1979distributed}. Also, if the initial prices are all zeros and $N\leq K$ the algorithm still converges in finite time and with the same bounds on optimality. The original Bertsekas auction algorithm is depicted in Table \ref{table_auction_alg}.
In \cite{yang2009} an algorithm based on the auction method
for two resource allocation problems in uplink OFDMA networks was proposed. Both single-cell and multi-cell scenarios were considered. In the single-cell scenario the auction algorithm was applied in a centralized manner at the base station. In the multi-cell scenario each cell determines the assignment for its users using the auction algorithm and sends the prices to its neighboring stations. This process continues until multi-cell optimality is achieved.
Note that each person raises his bid by the difference between his two most profitable objects. For that to be done, each user must know the prices of the objects and the other user bids on each iteration. In scenarios where no coordination between people is possible, they have no way of knowing the other person's bids or the prices of the objects. Hence, fully distributed methods are needed to solve the assignment problem when there is no knowledge of the other person's bids.

\section{The distributed auction algorithm}
The assumption in the original auction algorithm was that all the people involved know the price of each object; i.e, the highest bid.
This assumption does not hold when designing a distributed algorithm with no explicit communication between people.
In this section we propose a fully distributed version of the auction algorithm which does not require any explicit message passing or a shared memory. Like the auction algorithm, the distributed auction algorithm has two stages, the bidding stage and the assignment stage. The distributed auction algorithm can described as follows:
Let $\textbf{R}$ be an $N\times K$ nonnegative reward matrix and let $\textbf{B}$ be an $N\times K$ bid matrix. In the initialization stage each person sets his bids to be $0$; i.e., $\textbf{B}(n,k)=0,\forall n,k$, select $\epsilon>0$ and sets his state to unassigned. As in the original auction algorithm, the distributed auction algorithm is an iterative algorithm where each iteration is composed of a bidding stage where people raise their bids and an assignment stage where objects are assigned to the highest bidders. In the bidding stage, each unassigned person $n$ find his most profitable object $\tilde{k}_n$ and the profits from his most profitable object $\gamma_n$ and his second most profitable object $\omega_{n}$
\beq
\bea{l}
\displaystyle \tilde{k}_n=\arg\max_k\left(\textbf{R}(n,k)-\textbf{B}(n,k)\right)\\
\displaystyle \gamma_n=\textbf{R}(n,\tilde{k}_n)-\textbf{B}(n,\tilde{k}_n)\\
\displaystyle \omega_{n}=\max_{k\neq \tilde{k}_n}(\textbf{R}(n,k)-\textbf{B}(n,k))
\ena
\eeq
Each unassigned person raises the bid on his most profitable object by
\beq
\gamma_n-\omega_{n}+\epsilon
\eeq
and then all the people bid on an object. If a person $n$ is assigned to object $k$ he continues to bid on that object without raising his bid. If a person $n$ is unassigned he bids on $\tilde{k}_n$ with the new bid $\textbf{B}(n,\tilde{k}_n)$. In the assignment stage each object is assigned to the highest bidder. An object without bids stays unassigned and people who were not assigned to objects become unassigned. The bidding and assignment stages proceed in iterations until all the people are assigned to objects. Once all of the people are assigned to objects, no one raises his bid and as a result the assignment becomes static. When all the people are assigned we say that the algorithm has \emph{converged}. Note, that unlike the original auction algorithm, in the distributed auction algorithm the prices of the objects $\rho_k$ are not known to the bidders. Hence, the prices of the objects are determined locally by each bidder. In the distributed scenario we replace $\rho_{k}$ by $\textbf{B}(n,k)$ which is the price that the $n$'th person is willing to pay for the $k$'th object.
As a result, in the distributed auction algorithm the $\epsilon$-CS condition is replaced by a local condition. We call this modified condition Local $\epsilon$-complementary slackness (L$\epsilon$-CS).
The price that the $n$'th person pays on assignment $\boldsymbol\eta$ is denoted by $ \textbf{B}(n,\eta_n)$
The L$\epsilon$-CS condition with respect to an assignment $\boldsymbol\eta$ and a bidding matrix $\textbf{B} $ is defined by:
\beq
\textbf{R}(n,\eta_n)- \textbf{B}(n,\eta_n)\geq \max_k(\textbf{R}(n,k)-\textbf{B}(n,k))-\epsilon
\eeq
%where $\rho_{nk_n}$ is the highest bid the $n$'th user has made for object $k$ and
%where $\rho_{nk}$ is the price that a person $n$ is currently willing to pay for object $k$ and $\rho_{nk}$ is the price person $n$ is currently willing to pay for object $k_n$ which he is currently assigned to.
This condition relies solely on local information and hence is suitable for distributed algorithms.
We will now prove that if the L$\epsilon$-CS condition is satisfied for all the people, $\epsilon$-CS is also satisfied for all the people.
\theorem\label{theorem1} Given a feasible assignment $\boldsymbol\eta$, a reward matrix $\textbf{R}$ and a bidding matrix $\textbf{B}$ which satisfies L$\epsilon$-CS for all people then the $\epsilon$-CS condition is also satisfied for all the people.

\proof

If L$\epsilon$-CS condition is fulfilled, then
\beq
\label{local_cs}
	\textbf{R}(n,\eta_n)-\textbf{B}(n,\eta_n)\geq \max_k(\textbf{R}(n,k)-\textbf{B}(n,k))-\epsilon
\eeq
Since $\textbf{B}(n,k)\leq \max_n(\textbf{B}(n,k))$ it implies that
\beq
\bea{l}
\displaystyle\max_k(\textbf{R}(n,k)-\textbf{B}(n,k))\geq\max_k(\textbf{R}(n,k)-\max_n(\textbf{B}(n,k)))=\\
\displaystyle=\max_k(\textbf{R}(n,k)-\rho_k)
\ena
\eeq
where $\rho_k$ is defined in (\ref{eq_rho}).
Therefore
\beq
\label{eq7}
\max_k(\textbf{R}(n,k)-\textbf{B}(n,k))-\epsilon\geq\max_k(\textbf{R}(n,k)-\rho_k)-\epsilon
\eeq
On the other hand, since the assignment stage is the same between the algorithms, an object is assigned to the highest bidder. Hence, if an object $k$ is assigned to person $n$ it implies that
\beq
 \textbf{B}(n,\eta_n)=\max_{n}\textbf{B}(n,k)=\rho_{\eta_n}
\eeq
Hence
\beq
\label{eq_ls}
\textbf{R}(n,\eta_n)-\textbf{B}(n,\eta_n)=\textbf{R}(n,\eta_n)-\rho_{\eta_n}
\eeq
Substituting (\ref{eq_ls}),(\ref{eq7}) into (\ref{local_cs}) gives us
\beq
	\textbf{R}(n,\eta_n)-\rho_{\eta_n}\geq\max_k(\textbf{R}(n,k)-\rho_k)-\epsilon
\eeq
which is the complementary slackness condition (\ref{eq_ecs}).
This means that any conclusion regarding $\epsilon$-CS also applies to L$\epsilon$-CS. On particular, an assignment satisfying  L$\epsilon$-CS is $N\epsilon$ optimal.
\theorem
The distributed auction algorithm converges to a static assignment and a static set of prices within a finite number of iterations.
\label{theorem_finite_time1}

\proof
The proof requires an adaption of \cite{bertsekas1992auction} and is given in appendix \ref{appendix_proof1}.

The distributed auction algorithm appears in Table \ref{table_auction_alg}.
\begin{table*}
\caption{Auction algorithm and distributed auction algorithm}
\label{table_auction_alg}
\begin{minipage}[b]{0.45\linewidth}\centering
%\begin{table}
%\caption{Distributed auction algorithm}
\begin{center}
\textbf{Auction algorithm}
\end{center}
\begin{tabular}{l}
\hline
Select $\epsilon>0$, set all the people as unassigned and set\\
 $\rho_k=0,n=1..N$\\
\textbf{Repeat}\\
\ \ 1. Choose an unassigned person $n$\\
\ \ 2. Calculate his maximum profit $\gamma_{n}=\max_k(\textbf{R}(n,k)-\rho_k)$ \\
\ \ 3. Calculate the second maximum profit\\
\ \ \ \ \ $\tilde{k}=\arg\max_k(\textbf{R}(n,k)-\rho_k)$\\
\ \ \ \ \ $\omega_{n}=\max_{k\neq \tilde{k}}(\textbf{R}(n,k)-\rho_k)$\\
\ \ 4. Assign object $\tilde{k}$ to person $n$. If this object has been\\
\ \ \ \ \ assigned to another person, make this \\
\ \ \ \ \ person unassigned (and as a result unassigned).\\
\ \ 5. Set person $n$ as assigned\\
\ \ 6. Update the price of object $\tilde{k}$ to be \\
\ \ \ \ \ $\rho_{\tilde{k}}=\rho_{\tilde{k}}+\gamma_{n}-\omega_{n}+\epsilon$\\
\textbf{Until} all the people are assigned\\
\hline
\end{tabular}
\end{minipage}
\hspace{0.5cm}
\begin{minipage}[b]{0.45\linewidth}
%\begin{table}
%\caption{Auction algorithm }
\begin{center}
\textbf{Distributed auction algorithm}
\end{center}
\begin{tabular}{l}
\hline
Select $\epsilon>0$, set all the people as unassigned and set\\
$\textbf{B}(n,k)=0,\forall n,k$\\
\textbf{Repeat}\\
\ \ 1. Each unassigned person $n$ calculates its own maximum profit:\\
\ \ \ \ \ $\gamma_{n}=\max_k(\textbf{R}(n,k)-\textbf{B}(n,k))$\\
\ \ 2.  Each unassigned person $n$ calculates its second maximum profit: \\
\ \ \ \ \ $\tilde{k}_n=\arg\max_k(\textbf{R}(n,k)-\textbf{B}(n,k))$ \\
\ \ \ \ \ $\omega_{n}=\max_{k\neq \tilde{k}_n}(\textbf{R}(n,k)-\textbf{B}(n,k))$\\
\ \ 3. Each unassigned user $n$ updates the price of his best object $\tilde{k}_n$ \\
\ \ \ \ \ to be $\textbf{B}(n,\tilde{k}_n)=\textbf{B}(n,\tilde{k}_n)+\gamma_{n}-\omega_{n}+\epsilon$\\
\ \ 4. All the users bid. The unassigned people bid on their new best\\
\ \ \ \ \  object with the updated bid. The assigned users bid on the last\\
\ \ \ \ \  object they bid on and with the same price.\\
\ \ 5. Assign objects to the highest bidder (objects with no bids \\
\ \ \ \ \ stay unassigned)\\
\textbf{Until} all users are assigned\\
\hline
\end{tabular}
%\label{table_dist_auction_alg}
%\end{table}

%\label{table_auction_alg}
%\end{table}
\end{minipage}
\end{table*}
\subsection {Differences between the original auction algorithm and the distributed auction algorithm}
The main difference between the auction algorithm and the distributed auction algorithm is the information used to make a bid. In the auction algorithm each unassigned person needs to know all the other people's bids in order to make his bid. In the distributed auction algorithm each unassigned person relies solely on the local information available to determine his bid. More specifically, in the original auction algorithm, after each assignment phase each person knows the maximal bid on all the objects. As a result if a person was unassigned in the $i$'th iteration his next bid would be high enough so that he would be assigned on iterations $i+1$. However, in the distributed auction algorithm, after each assignment phase each bidder only knows if he was assigned or not. If he got assigned he was the highest bidder and there is no need to raise his bid. If he did not get assigned the bidder knows that he was not the highest bidder and he needs to raise the bid on his most profitable object. However, since each unassigned person in the distributed auction algorithm raised his bid solely according to his past bids there is no guarantee that a person who was unassigned in the $i$'th iteration would get assigned in iteration $i+1$ in the distributed auction algorithm. Furthermore, the bidding results might differ between the algorithms. However, Theorem \ref{theorem1} guarantees the convergence of the distributed auction algorithm to a solution that is within $N\epsilon$ from the optimal solution and Theorem \ref{theorem_finite_time1} guarantees the convergence of the distributed auction algorithm in finite time.

Up to this point we assumed $N\leq K$. However, similar to the algorithm in \cite{bertsekas1991reverse}, the algorithm also converges when $N>K$. The case of $N> K$ is equivalent to the case of $N=K$ with $N-K$ zero columns in $\textbf{R}$ \cite{bertsekas1991reverse}. In that case if there exists a person $n$ where
\beq
\textbf{R}(n,k)-\textbf{B}(n,k)<0, \forall k\leq N
\eeq
then
\beq
\max_k(\textbf{R}(n,k)-\textbf{B}(n,k))=0
\eeq
this implies
\beq
\textbf{R}(n,\eta_n)=0
\eeq
which is equivalent to not being assigned at all. Hence, if the profit from all the objects is negative the person stops bidding and sets his state to assigned. Because of the equivalence between the $N>K$ case and the $N=K$ with zero columns case, the algorithm will converge to a solution within $N\epsilon$ from the optimal solution.

%%%%%%%%%%%%%%%%%%%%%%%%%%%%%%%%%%להוסיף דיון בהתכנסות ועצירה למקרים שונים של N,K
%Another difference is that because of its distributed nature, the termination mechanism also relies on local observations.
%From \cite{bertsekas1979distributed} we know that in the case of $N=K$, the algorithm terminates once all objects have %bids.
%We can use this observation to define a termination criterion based on local observation.
%Each user can keep a list of all the channels that have been used previously. When the list has $K$ channels the algorithm %terminates.
%In the case where $N\neq K$ this is not true and the stopping criterion is not valid.
%Thus there is no termination criterion for the algorithm in the case of $N\neq K$. Once all the users are assigned, the %prices (and as a result the assignments) stay the same because no user updates his prices.

\section {Applying the distributed auction algorithm using Opportunistic CSMA}
\label{section_auction_csma}
\begin{figure}[htbp]
\centering \includegraphics[width=0.5\textwidth]{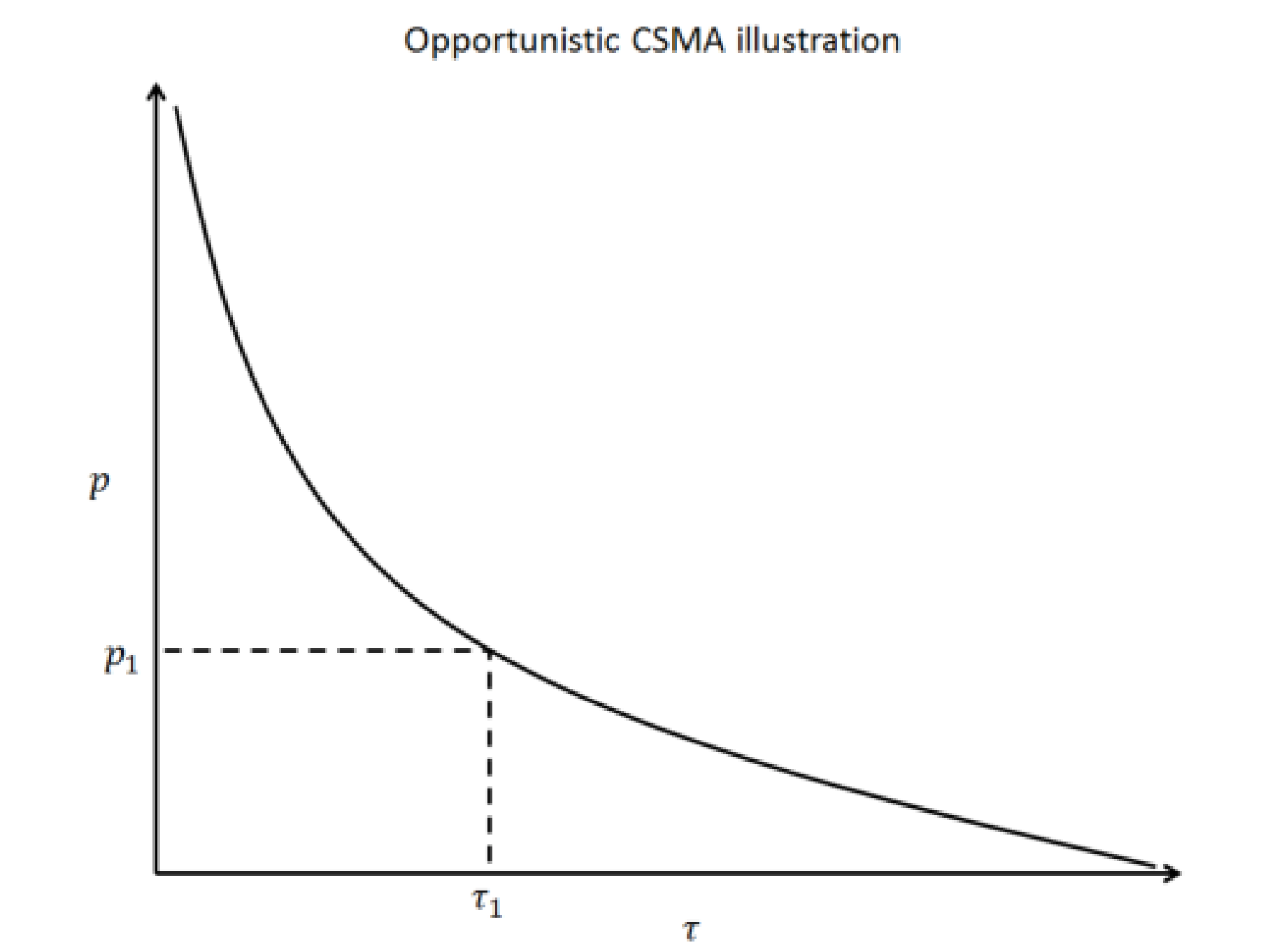}
\caption{An example of a function that maps better fitness to shorter time}
\label{figure_opp_csma}
\end{figure}

In this section we show how to implement the modified auction algorithm in a fully distributed manner using opportunistic CSMA.
Opportunistic CSMA \cite{zhao2005opportunistic} is a distributed transmission protocol suggested for wireless sensor networks. The protocol is made up of carrier sensing and a backoff strategy. Continuous sensing of all channels by all users is assumed. Each user in the network calculates a fitness measure $\psi_n$ and maps it to a backoff time $\tau_n$ based on a predetermined common decreasing function $f(\psi_n)$. Figure \ref{figure_opp_csma} shows an example of such a backoff function.
 Each user listens to the channel he wants to use and if no other user transmits before its backoff time expires, the user is allowed to transmit. This protocol results in the user with highest $\psi_n$ transmitting in the channel. In other words, this protocol can be used to determine the maximum of a vector in a distributed manner. In the distributed auction algorithm the only coordination needed between users is an auctioneer to decide which user made the highest bid. The opportunistic CSMA can be used as an auctioneer.
Since the opportunistic CSMA protocol does not require any message passing between users, it can be used to implement a fully distributed solution on the maximum sum-rate problem. The resulting assignment is within $N\epsilon$ from the optimal solution as in the auction algorithm. We can define the reward that each user $n$ gets from channel $k$ to be the achievable rate on that channel $\textbf{R}(n,k)$.
Using the opportunistic CSMA scheme, each user $n$ tries to access his best profit channel defined by
\beq
\tilde{k}_n=\arg\max_k(\textbf{R}(n,k)-\textbf{B}(n,k))
\eeq
with a backoff time of
\beq
\tau_{n}=f(\rho_{n,\tilde{k}_n})
\eeq
where $f(x)$ is a positive monotonically decreasing function. The price $\textbf{B}(n,k)$ is determined and updated if necessary as described in section \ref{section_auction_alg}.
The prices and their corresponding waiting times must converge in a finite number of iterations as in the distributed auction algorithm.
The distributed auction algorithm for cognitive radio systems using opportunistic CSMA is described in Table \ref{table_dist_auction_CSMA}.
\begin{table}
\caption{Distributed auction algorithm using opportunistic carrier sensing}
\begin{tabular}{l}
\hline
Select $\epsilon>0$, each user sets himself as unassigned\\
and set $\textbf{B}(n,k)=0,\forall k$\\
\textbf{Repeat}\\
\ \ 1. Each unassigned user $n$ calculates its own maximum profit:\\
\ \ \ \ \ $\gamma_{n}=\max_k(\textbf{R}(n,k)-\textbf{B}(n,k))$\\
\ \ 2. Each unassigned user $n$ calculates its own second maximum profit: \\
\ \ \ \ \ $\tilde{k}_n=\arg\max_k(\textbf{R}(n,k)-\textbf{B}(n,k))$ \\
\ \ \ \ \ $\omega_{n}=\max_{k\neq \tilde{k}_n}(\textbf{R}(n,k)-\textbf{B}(n,k))$\\
\ \ 3. Each unassigned user $n$ updates the price of his best channel $\tilde{k}_n$ \\
\ \ \ \ \ to be $\textbf{B}(n,\tilde{k}_n)=\textbf{B}(n,\tilde{k}_n)+\gamma_{n}-\omega_{n}+\epsilon$\\
\ \ 4. Each unassigned user $n$ maps the new best profit into \\
\ \ \ \ \ backoff time $\tau_{n}=f(\textbf{B}(n,\tilde{k}_n))$\\
\ \ 5. Each user waits for $\tau_{n}$ milliseconds. If the user detects that\\
\ \ \ \ \ $\tilde{k}_n$ is free when its backoff time has expired, he\\
\ \ \ \ \  transmits his packet on $\tilde{k}_n$ and sets his state to assigned.\\
\textbf{Until} all users are assigned\\
\hline
\end{tabular}
\label{table_dist_auction_CSMA}
\end{table}

Note that if the users are capable of measuring the back-off times of the other users, the original auction algorithm can be applied.
\section{Upper bound on the expected number of iterations in the distributed auction algorithm}
\label{section_ub_iter}
In Theorem \ref{theorem_finite_time1} we proved that the distributed auction algorithm converges in finite time. In this section we give an upper bound on the expected number of iterations required for the algorithm to converge to the optimal solution. We will first prove the following simple lemma:

\lemma\label{lemma_ub} Let $\bf{R}$ be an $N\times N$ nonnegative matrix. Let $T$ be the number of iterations until the distributed auction algorithm converges and let $T_n$ be the number of iterations in which user $n$ was unassigned; then
\beq
T_n\leq K+\frac{1}{\epsilon}\sum_{k=1}^K\textbf{R}(n,k)
\eeq

\proof
In each iteration in which the $n$'th user is unassigned he reduces his profit on his most profitable channel  by at least $\epsilon$. Hence, if
\beq
T_n= K+\frac{1}{\epsilon}\sum_{k=1}^K\textbf{R}(n,k)
\eeq
then
\beq
\textbf{R}(n,k)-\textbf{B}(n,k)<0,\forall k
\eeq
and therefore
\beq
\textbf{B}(n,k)>0,\forall k
\eeq
which by Theorem \ref{theorem_finite_time1} implies
\beq
T\leq T_n
\eeq
and as a result
\beq
T_n\leq K+\frac{1}{\epsilon}\sum_{k=1}^K\textbf{R}(n,k)
\eeq
%\begin{flushright}
%$_{_{\blacksquare}}\ \ \ \ $
%\end{flushright}
\theorem\label{theorem_ub1} Let $\bf{R}$ be an $N\times N$  i.i.d random matrix. Let $r>0$ be a nonnegative random variable with a Probability Density Function (PDF) $f_r(r_0)$ and assume for all $n,k$
\beq
f_{\textbf{R}(n,k)}(r_0)=f_{r}(r_0),\forall n,k
\eeq
Let $T$ be the number of iterations until the distributed auction algorithm converges;
then:
\beq
E\left(T\right)\leq  \frac{1}{\epsilon}N^2E(r)+N^2
\eeq

\proof

The time it takes for the distributed auction algorithm until all users are assigned is bounded by:
\beq
T\leq\sum_{n=1}^N T_n
\eeq
and by Lemma \ref{lemma_ub} we get
\beq
T\leq\sum_{n=1}^N K+\frac{1}{\epsilon}\sum_{k=1}^K\textbf{R}(n,k)
\eeq
Taking the expected value gives us
\beq
E(T)\leq NK+\frac{1}{\epsilon}KNE(r)
\eeq
and since we assumed $K=N$
\beq
E(T)\leq N^2+\frac{1}{\epsilon}N^2E(r)=O\left(\frac{1}{\epsilon}N^2E(r)\right)
\eeq

Note that as shown in \cite{bertsekas1979distributed}, if $\textbf{R}$ is integer valued and $\epsilon<\frac{1}{N}$ then the auction algorithm converges to the optimal solution. Hence if \textbf{R} is a quantized matrix with quantization $\frac{1}{q}$ the choice of $\epsilon<\frac{1}{qN}$ is sufficient to obtain the optimal solution.

\section{Asymptotically optimal auction algorithm with reduced complexity}
\label{section_performance}
As shown in the previous section, the expected time complexity of the distributed auction algorithm is bounded by $O\left(\frac{1}{\epsilon}N^2E(r)\right)$. This complexity is too high for many practical implementations. However, in real life systems there is a limit on the maximal number of allowed bits per transmission. We exploit this fact to devise a variation of the distributed auction algorithm that has a reduced expected time complexity. We call this variant "the truncated auction algorithm" (TAA). The TAA has a reduced worst case time complexity and is asymptotically optimal under the assumption of bounded maximal transmission rate per channel use. % The truncated auction algorithm use the distributed auction algorithm with different choice of the utility matrix.
%In the feasibility auction algorithm the rate matrix is replaced with a binary $\{0,1\}$ matrix where the best $\alpha\log(N), %\alpha>2$ channels of each user are replaced by $1$ and the other entries are replaced by $0$. Applying the distributed auction %algorithm on this utility matrix is equivalent to finding a feasible assignment (not necessarily optimal) using only the best %$\alpha\log(N)$ channels of each user. We show that the worst case complexity of this algorithm is $O\left(\alpha N^2\log(N)\right)$ %and that this algorithm is asymptotically sum-rate optimal under the assumption of bounded maximal rate per channel use.
%In the truncated auction algorithm %we replace the utility matrix with a truncated version of the original matrix.
%By truncated we mean that we keep the rates of the best $\alpha\log_2(N), \alpha>1$ channels as in the original rate matrix and replace all the other channel rates by $0$. The distributed auction algorithm is then applied on the revised rate matrix. We show that the probability of not converging to the optimal assignment using the truncated rate matrix is decreasing as $\frac{1}{N^{\alpha-1}}$ and the expected time complexity of this algorithm is bounded by $O\left(N^2\alpha\log_2(N)E(X)\right)$ for integer valued matrices.
The TAA is the same as the distributed auction algorithm with a simple change of the rate matrix. In this variation of the distributed auction algorithm each user maintains the rates of his best $\alpha\log_2(N)$ channels in the rate matrix and replaces the rest of the rates in the rate matrix with zeros. The distributed auction algorithm is applied on the revised rate matrix. Note that our simulations show that the expected convergence times of the distributed auction algorithm and the TAA are the same. However, we can prove this for the TAA. The algorithm is depicted in Table \ref{table_truncated_auction}.
\begin{table}
\caption{Truncated auction algorithm}
\begin{tabular}{l}
\hline
1.Select $\epsilon=\frac{1}{N}$,each user sets himself as unassigned\\
2.Each user sets the prices for all channels\\
 to zero $\textbf{B}(n,k)=0,n=1..N,k=1..K$\\
3. Each user sets all the rates other than the best\\
\ \ $\alpha\log_2(N)$ best channels to $0$\\
\textbf{At the beginning of each time slot}\\
\ \ 1. Each unassigned person $n$ calculates its own maximum profit\\
\ \ \ \ \ $\gamma_{n}=\max_k(\textbf{R}(n,k)-\textbf{B}(n,k))$ \\
\ \ 2. Each unassigned person $n$ calculates its own second maximum profit \\
\ \ \ \ \ $\tilde{k}_n=\arg\max_k(\textbf{R}(n,k)-\textbf{B}(n,k))$ \\
\ \ \ \ \ $\omega_{n}=\max_{k\neq \tilde{k}_n}(\textbf{R}(n,k)-\textbf{B}(n,k))$\\
\ \ 3. Each unassigned user $n$ updates the price of his best object $\tilde{k}_n$ \\
\ \ \ \ \ to be $\textbf{B}(n,\tilde{k}_n)=\textbf{B}(n,\tilde{k}_n)+\gamma_{n}-\omega_{n}+\epsilon$\\
\ \ 4. Each unassigned user $n$ maps the new best profit into \\
\ \ \ \ \ back-off time $\tau_{n}=f(\textbf{B}(n,\tilde{k}_n))$\\
\ \ 5. Each user waits for $\tau_{n}$ milliseconds. If the user detects that\\
\ \ \ \ \ $\tilde{k}_n$ is free when its back-off time has expired, he\\
\ \ \ \ \  transmits his packet on the channel and sets his state to assigned.\\
\hline
\end{tabular}
\label{table_truncated_auction}
\end{table}
The main result of this section is as follows:
\theorem Let \textbf{R} be an $N\times N$ i.i.d random matrix. Let $r\in[0,a]$ be a nonnegative random variable with PDF $f_{r}(a)>0$ and assume that for all $n,k$
 \beq
 f_{\textbf{R}(n,k)}(r_0)=f_{r}(r_0),\forall n,k
 \eeq
 Let $R_{opt}$ be the sum-rate achieved by the optimal assignment and let $R_{trunc}$ be the sum-rate achieved by the TAA using only the best $\alpha\log_2(N)$ channels of each user where $\alpha>1$.
 Then:
\begin{enumerate}
\renewcommand{\theenumi}{\alph{enumi}}
\label{theorem_truncated}
\item \label{theorem_truncated_1} The expected time complexity of the TAA is bounded by $O\left(\frac{\alpha}{\epsilon}N\log_2(N)E(r)\right)$
\item \label{theorem_truncated_2} $E\left(R_{trunc}\right)\geq\left(1-\frac{1}{N^{\alpha-1}}\right)E\left(R_{opt}\right)$
\item \label{theorem_truncated_3} $\lim_{N\to \infty}E(R_{opt})-E(R_{trunc})=0$
\end{enumerate}
\proof

We first prove \ref{theorem_truncated}-\ref{theorem_truncated_1}. The proof is almost the same as the proof to the bound on the expected time complexity of the distributed auction algorithm. In the TAA there are $\alpha N\log_2(N)$ nonzero entries in the rate matrix instead of $N^2$ nonzero entries in the original distributed auction algorithm.
As a result, in the TAA using lemma \ref{lemma_ub} we get
\beq
E(T)\leq \sum_{n=1}^NE(T_n)\leq N^2+\frac{\alpha}{\epsilon}N\log_2(N)E(r)
\eeq
 This concludes the proof of \ref{theorem_truncated}.\ref{theorem_truncated_1}.

To prove \ref{theorem_truncated}-\ref{theorem_truncated_2} we first show that the probability that the assignment is obtained by the TAA using only the best $\alpha\log_2(N)$ channels of each user is optimal with a probability greater than $1-\frac{1}{N^{\alpha-1}}$. To prove \ref{theorem_truncated}-\ref{theorem_truncated_2} we use the following theorem proven in \cite{aldous2001zeta}:
 \theorem \label{theorem_aldous} Let $\textbf{A}$ be an $N\times N$ i.i.d matrix with a common random variable $X$ and PDF $f_X(x)$. If $f_X(0)>0$, the probability that user $i$ will get assigned to his $k$'th best channel in the minimum assignment problem approaches $2^{-k}$ for large $N$.

It also follows from Theorem \ref{theorem_aldous} that if $X\in[a,b]$ and $f_X(b)>0$ the same probabilities hold for the maximum sum assignment problem since we can define a new random variable $Y=b-X$ where the minimum assignment for $Y$ is the maximum assignment for $X$.
The probability that a user will get assigned to one of his best $k$ channels is
\beq
P(c_i\leq k)=1-2^{-k}
\eeq
where $c_i$ is the rank of the assigned value to $i$'th user.
The probability that all users were assigned to one of their best $k$ channels in the optimal assignment is bounded by
\beq
\label{eq_union1}
\bea{l}
\displaystyle P\left(\cap_{i=1}^NP\left(c_i\leq k\right)\right)\geq P(c_1\leq k)-\sum_{i=2}^N2^{-k}=\\
=1-N2^{-k}
\ena
\eeq
If we choose $k=\alpha\log_2(N)$ the probability that each user will be assigned to one of his best $k$ channels becomes:
\beq
P\left(\cap_{i=1}^NP\left(c_i\leq k\right)\right)\geq1-\frac{1}{N^{\alpha-1}}
\eeq
This implies that the TAA will converge to the optimal assignment with a probability greater than $1-\frac{1}{N^{\alpha-1}}$. Now, if the truncated auction algorithm does not converge to the optimal solution we assume that the resulting sum-rate is zero.
The above gives us the desired result
\beq
E\left(R_{trunc}\right)\geq\left(1-\frac{1}{N^{\alpha-1}}\right)E\left(R_{opt}\right)
\eeq
This concludes the proof of part \ref{theorem_truncated}-\ref{theorem_truncated_2}.

Part \ref{theorem_truncated}-\ref{theorem_truncated_3} is trivially obtained from part \ref{theorem_truncated}-\ref{theorem_truncated_2} by letting $N$ go to infinity.
\section{Asymptotically Optimal greedy assignment in Rayleigh fading channels}
\label{section_opt_rayleigh}
In this section we show that in the case of i.i.d Rayleigh fading channels a simple randomized greedy assignment scheme is asymptotically optimal. We also give an upper bound on the expected rate loss per user using the greedy scheme.
The randomized greedy assignment is obtained by randomly choosing unassigned users one at a time and assigning them to their best channel out of the remaining unassigned channels. The process is repeated until all the users are assigned.
The randomized greedy assignment scheme is shown in Table \ref{table_gca}.
\begin{table}
\caption{Randomized greedy channel assignment}
\begin{enumerate}
\item Set all users and channels as unassigned.
\item While there are unassigned users do:
    \begin{enumerate}
    \item Select a random user out of the unassigned users
    \item Assign the selected user to his best unassigned channel
    \item Set the user and channel as assigned.
    \end{enumerate}
\item Return to 2.
\item End
\end{enumerate}
\label{table_gca}
\end{table}

In this section we assume each channel is a Rayleigh fading channel; i.e., the channel attenuation $|h_n(k)|^2$ is an exponential random variable given by:
\beq
|h_n(k)|^2=G_n\cdot F_n\cdot \frac{1}{r_n^{\alpha}}
\eeq
where $G_n$ is a global normalizing factor, $F_n$ is an exponentially distributed gain (due to the
Rayleigh fading channel with a multipath effect) and $\alpha$ is the path loss exponent; therefore:
\beq
f_{|h_n(k)|^2}(x)= \frac{r_n^{\alpha}}{G_n}e^{-\frac{r_n^{\alpha}}{G_n}x}
\eeq
and
\beq
f_{\SNR_n(k)}(x)= \lambda_ne^{-\lambda_nx}
\eeq
where
\beq
\label{eq_lambda}
 \lambda_n=\frac{\sigma_n^2r_n^{\alpha}}{p_n\cdot G_n}
\eeq

The rate of user $n$ in channel $k$ is given by \cite{cover1991elements}:
\beq
\label{eq_rtae}
R_n(k)=\log\left(1+\SNR_n\left(k\right)\right)
\eeq
The cumulative distribution function of $R(k)$ is given by:
\beq
F_{R(k)}(y)=F_{\SNR}(e^{y}-1)=1-e^{-\lambda(e^{y}-1)}
\eeq

%since $R(k)$ is a positive random variable. Its expectation is given by \cite{ross2009introduction}:
%\beq
%\bea{l}
%\displaystyle E(R(k))=\int_0^{\infty}\left(1-F_{R(k)}(y)\right)dy=\\ \displaystyle %=\int_0^{\infty}\left(e^{-\lambda_n(e^{y}-1)}\right)dy= e^{\lambda_n}\int_0^{\infty}e^{-\lambda_n e^{y}}dy=\\
%\displaystyle =e^{\lambda_n}\int_{\lambda_n}^{\infty}\frac{e^{-t}}{t}dt=e^{\lambda_n}E_1(\lambda_n)
%\ena
%\eeq
Let $R_{l:K}$  be the $l$'th smallest random number out of $K$ random numbers.
The expected $R_{l:K}$ is given by \cite{alouini1999}:
\beq
\label{eq_ord_stats}
\bea{l}
E\left(R_{l:K}\right)=\\
\binom{K}{l}\sum_{m=1}^l\binom{l}{m}\frac{(-1)^{m+1}m}{K-l+m}e^{\lambda(K-l+m)}E_1(\lambda(K-l+m))
\ena
\eeq
where $E_1(\cdot)$ is the exponential integral defined by:
\beq
\displaystyle E_1(x)=\int_x^{\infty}\frac{e^{-t}}{t}dt
\eeq
The following lemmas are needed for the proof of the asymptotical optimality of the greedy scheme in i.i.d Rayleigh fading channels.

\lemma Assume $R$ is an $N\times N$ i.i.d matrix with random rates defined in equation (\ref{eq_rtae}). Then the expected sum-rate of the greedy assignment scheme in i.i.d Rayleigh fading channels is given by:
\beq
\label{eq_lb}
\displaystyle L= \sum_{m=K-N+1}^K \sum_{j=1}^m \binom{m}{j}(-1)^{j+1}e^{j\lambda}E_1(j\lambda)
\eeq
\proof
The $k$'th assigned user gets assigned to his best channel out of $N-k+1$ available channels unconditioned on the previous picks.
This is true because the users are picked in a random order.
The expected sum-rate of the greedy scheme is given by:
\beq
\label{eq_L1}
\sum_{m=K-N+1}^K R_{m:K}
\eeq
and by substituting (\ref{eq_ord_stats}) into (\ref{eq_L1}) we get
\beq
\sum_{m=K-N+1}^K \sum_{j=1}^m \binom{m}{j}(-1)^{j+1}e^{j\lambda}E_1(j\lambda)
\eeq
 \lemma Assume $R$ is an $N\times N$ i.i.d matrix with random rates defined in equation (\ref{eq_rtae}).
 An upper bound on the expected optimal assignment is given by:
\beq
\label{eq_ub}
\displaystyle U= N\sum_{m=1}^K\binom{K}{m}(-1)^{m+1}e^{m\lambda}E_1(m\lambda)
\eeq
\proof

The proof is trivial since it is the expected sum-rate when assuming each user was assigned to his best channel every time. This is of course an upper bound on the optimal assignment.

We now show that the greedy scheme is asymptotically optimal in both low and high $\SNR$ regimes.
\subsection{Asymptotic optimality in the low $\SNR$ regime}
In this section we analyze the performance of the greedy scheme in low $\SNR$ and show that it is asymptotically optimal for i.i.d Rayleigh channels.
 To analyze the performance of the greedy scheme in the low $\SNR$ regime we assume for simplicity that $N=K$.
 However, the analysis holds for any $N\leq K$.
\theorem Assume $R$ is an $N\times N$ i.i.d random matrix. Each element in the matrix is defined as in (\ref{eq_rtae})
and $L,U$ defined in (\ref{eq_lb}), (\ref{eq_ub}); then in the low $\SNR$ regime:
\beq
\lim_{\SNR \to 0}\left(\frac{U-L}{U}\right)\leq\frac{1}{H_N}-\frac{1}{N}
\eeq
 where $H_N$ is the $N$'th harmonic number defined by:
\beq
\displaystyle H_N=\sum_{k=1}^N\frac{1}{k}
\eeq
\proof

We use the following known bounds on $E_1(m\lambda)$ \cite{abramowitz1964handbook}
\beq
\bea{l}
\frac{1}{2}e^{-m\lambda}\log\left(1+\frac{2}{m\lambda}\right)<E_1(m\lambda)<\\<e^{-m\lambda}\log\left(1+\frac{1}{m\lambda}\right)
\ena
\eeq
 In low $\SNR$ $\frac{1}{m\lambda}<<1$  and
 \beq
 \frac{1}{m\lambda}-\frac{1}{2m^2\lambda^2}\leq\log(1+\frac{1}{m\lambda}) \leq \frac{1}{m\lambda}
 \eeq
 In that case the bounds are tight and we can bound $E_1(m\lambda)$ by
\beq
e^{-m\lambda}\left(\frac{1}{m\lambda}-\frac{1}{2m^2\lambda^2}\right) \leq E_1(m\lambda)\leq e^{-m\lambda}\frac{1}{m\lambda}
\eeq
Using \cite{flajolet1995mellin} the bounds can be bounded by:
 \beq
 \bea{l}
 \displaystyle U \leq N\sum_{m=1}^N\binom{N}{m}(-1)^{m+1}\frac{1}{m\lambda}=\frac{N}{\lambda}H_N\\
 \displaystyle L\geq \sum_{m=1}^N \sum_{j=1}^m \binom{m}{j}(-1)^{j+1}\left(\frac{1}{m\lambda}-\frac{1}{2m^2\lambda^2}\right)=\\
 \displaystyle=\sum_{m=1}^N\left(\frac{H_m}{\lambda}-\left(\frac{\zeta(2)-H^{(2)}_m}{4\lambda^2}\right)-\frac{\left(H_m\right)^2}{4\lambda^2}\right)%\geq\\
  %\displaystyle\geq \sum_{m=1}^N\left(\frac{H_m}{\lambda}-\left(\frac{\zeta(2)-H^{(2)}_m}{4\lambda^2}\right)-\frac{\left(H_m\right)^2}{4\lambda^2}\right)
 \ena
 \eeq
where $H^{(2)}_N$ is defined by:
\beq
\displaystyle H_N^{(2)}=\sum_{k=1}^N\frac{1}{k^2}
\eeq

We can now give an upper bound on the relative error of the bounds by computing the relative difference between the upper and lower bounds. The relative difference in low $\SNR$ is given by:
\beq
\bea{l}
\displaystyle \frac{U-L}{U}\leq \frac{\sum_{m=1}^NH_N-H_m}{NH_N}+\\
\displaystyle+\frac{1}{4\lambda N H_N}\sum_{m=1}^N\left(\frac{\pi^2}{6}-H_m^{(2)}+\left(H_m\right)^2\right)\leq \\
\displaystyle\leq\frac{1}{H_N}-\frac{1}{N}+\\
\displaystyle+\frac{1}{4\lambda}\left(H_N+\frac{\pi^2}{6 H_N} -\frac{(N+1) H_{N+1}^{(2)}-H_{N+1}}{N H_N}\ \right)\leq\\
\displaystyle \leq \frac{1}{H_N}-\frac{1}{N}+\frac{H_N}{\lambda}\left(\frac{1}{8}+\frac{\pi^2}{54}\right)
\ena
\eeq
and for any fixed $N$ when $\SNR\to 0$
\beq
\lim_{\lambda \to \infty}\left(\frac{U-L}{U}\right)\leq\frac{1}{H_N}-\frac{1}{N}
\eeq
%This implies that for an arbitrary number of users $N$, $\frac{U-L}{U}$ behaves like
%\beq
%\frac{U-L}{U}\leq \frac{c\log(N)}{\lambda}+\frac{1}{\log(N)} , c<1
%\eeq.
%Typically the first term is the dominant unless $\SNR<<\frac{1}{log(N)}$.
%For example, if $\SNR=-20$dB and the number of users $N\leq 350$ then the first term will contribute less than $2\%$ the bound and for $\SNR=-30$dB the first term will contribute less than $1\%$ for any $N\leq 10^{14}$. This means that if the $\SNR$ is sufficiently low then the first term of the bound wold not be the dominant term for any reasonable number of users $N$.
Furthermore, since $H_N$ is diverging, if $H_N=o(\lambda)$:
\beq
\lim_{N \to \infty}\left(\frac{U-L}{U}\right)=0
\eeq
This implies that the bounds are asymptotically tight in the low $\SNR$ regime as the number of users grows.
This also implies that the greedy scheme is asymptotically optimal as the number of users goes to infinity.
\subsection{Asymptotic optimality in the high $\SNR$ regime}
We now analyze the performance of the greedy scheme in the high $SNR$ regime.
\theorem Assume $R$ is an $N\times N$ i.i.d random matrix. Each element in the matrix is defined as in (\ref{eq_rtae})
and $L,U$ defined in (\ref{eq_lb}), (\ref{eq_ub}); then in the high $\SNR$ regime for any $N\geq2$ the following holds:
\begin{enumerate}
\renewcommand{\theenumi}{\alph{enumi}}
\item
\beq
\bea{l}
\displaystyle \lim_{\lambda\to 0 }\left(U-L\right)=c\leq \\
\displaystyle \leq \left(N-1\right)\log\log(N-1)+\gamma +\frac{1}{\log (N)}+\\
\displaystyle +\left(N-2\right)\left(\gamma\left( \frac{1}{log(N)}-\frac{1}{log(2)}\right)-\log\log(2)\right)
\ena
\eeq

\item $\lim_{\SNR \to \infty} \frac{U-L}{U}=0$
\end{enumerate}

\proof

The series representation of $E_1(m\lambda)$ is given by:
\beq
\displaystyle E_1(m\lambda)=-\gamma -\log(m\lambda)-\sum_{l=1}^{\infty}\frac{(-m\lambda)^l}{ll!}
\eeq
where $\gamma$ is Euler's constant.
In the high $\SNR$ case, $m\lambda<<1$ and we can bound $E_1(m\lambda)$  by
\beq
\displaystyle -\gamma -\log(m\lambda)\leq E_1(m\lambda)\leq -\gamma -\log(m\lambda)+m\lambda
\eeq
Using these bounds we can now show that the difference between the higher and lower bound is asymptotically constant.
The bounds in the high $\SNR$ are approximately:
\beq
\bea{l}
\displaystyle U\leq -N(\gamma +log(\lambda))\left(1-\left(1-e^{\lambda}\right)^N\right)+\\ \displaystyle +N\sum_{m=1}^N\binom{N}{m}(-1)^{m}e^{m\lambda}\log(m)+\\
\displaystyle+N^2\lambda e^{\lambda}\left(1-e^{\lambda}\right)^{N-1}\\
\displaystyle L\geq -(\gamma +log(\lambda))\sum_{m=1}^N \left(1-\left(1-e^{\lambda}\right)^m\right)+\\ \displaystyle + \sum_{m=1}^N \sum_{j=1}^m \binom{m}{j}(-1)^{j}e^{j\lambda}\log(m)
\ena
\eeq
The difference between the higher and the lower bound is bounded by:
\beq
\bea{l}
\displaystyle U-L\leq  -N\left(\gamma +log\left(\lambda\right)\right) \left(1-e^{\lambda}\right)+\\
\displaystyle +N\sum_{m=1}^N\binom{N}{m}(-1)^{m}e^{m\lambda}\log(m)+\\
\displaystyle +N^2\lambda e^{\lambda}\left(1-e^{\lambda}\right)^{N-1}-\\
\displaystyle- \sum_{m=1}^N \sum_{j=1}^m \binom{m}{j}(-1)^{j}e^{j\lambda}\log(j)
\ena
\eeq

and as the $\SNR$ goes to infinity, using the results from \cite{flajolet1995mellin} we obtain:
 \beq
 \bea{l}
 \displaystyle \lim_{\lambda \to 0}\left( U-L\right)\leq N\sum_{m=1}^N\binom{N}{m}(-1)^{m}\log(m)-\\
  \displaystyle -\sum_{m=1}^N \sum_{j=1}^m \binom{m}{j}(-1)^{j}\log(j)=c\leq\\
  \displaystyle \leq \left(N-1\right)\log\log(N-1)+\gamma +\frac{1}{\log (N)}+\\
  \displaystyle +\left(N-2\right)\left(-\log\log(2)+\gamma\left( \frac{1}{log(N)}-\frac{1}{log(2)}\right)\right)
 \ena
 \label{eq_const}
 \eeq
 $c$ is a constant independent of $\lambda$. This shows that the bounds are tight in the high $\SNR$ regime as well since:
  \beq
 \displaystyle \lim_{\SNR \to \infty} \frac{c}{U}=0
 \eeq
This implies that the greedy scheme is asymptotically optimal in the high $\SNR$ regime regardless of the number of users.
Note that the expression in (\ref{eq_const}) could be used to give an estimate on the total or the relative error of the bounds in any $\SNR$.

\section{Simulations}
\label{section_simulations}
In this section we report some simulated results of the algorithms suggested in this paper.
We first show the average sum-rate achieved by the distributed auction algorithm and compare the results with the auction algorithm where the bids are known to all users.
We consider a system with $m=10$ users and $k=10$ channels.
The rates $\textbf{R}(n,k)$ are random rates in Rayleigh fading channels $\SNR=20db$.
\begin{figure}[htbp]
\centering \includegraphics[width=0.5\textwidth]{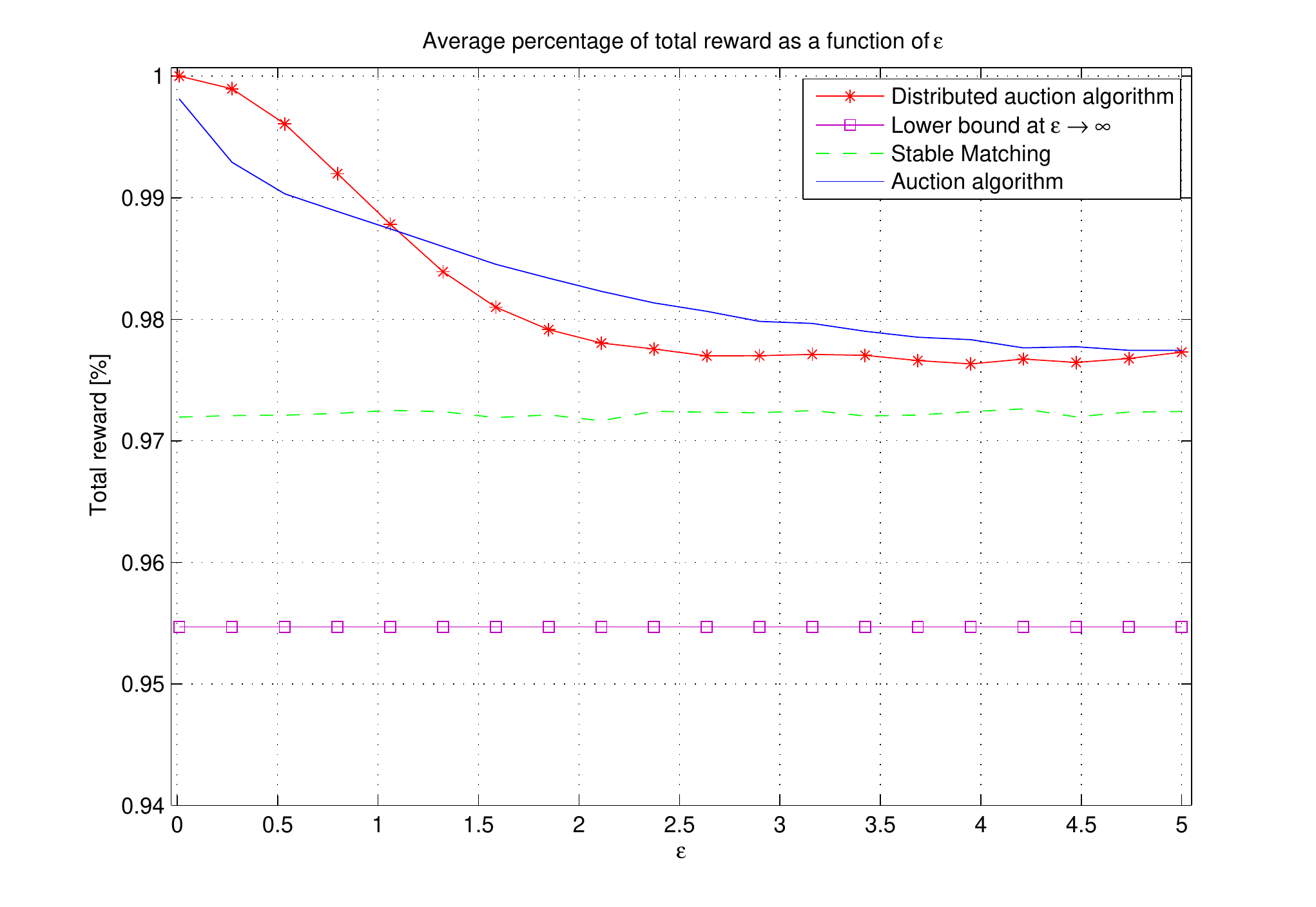}
\caption{Average distance from optimality as a function of $\epsilon$}
\label{figure_opt_comp}
\end{figure}
In Fig. \ref{figure_opt_comp} we show a comparison between the average results of both algorithms at termination as a function of $\epsilon$. It is easy to see that both algorithms are within $M\epsilon$ of being optimal. As $\epsilon$ gets larger, both
algorithms converge to a value that is better than the stable matching greedy solution. However, the degradation of the distributed algorithm is faster as a function of $\epsilon$.

\begin{figure}[htbp]
\centering \includegraphics[width=0.5\textwidth]{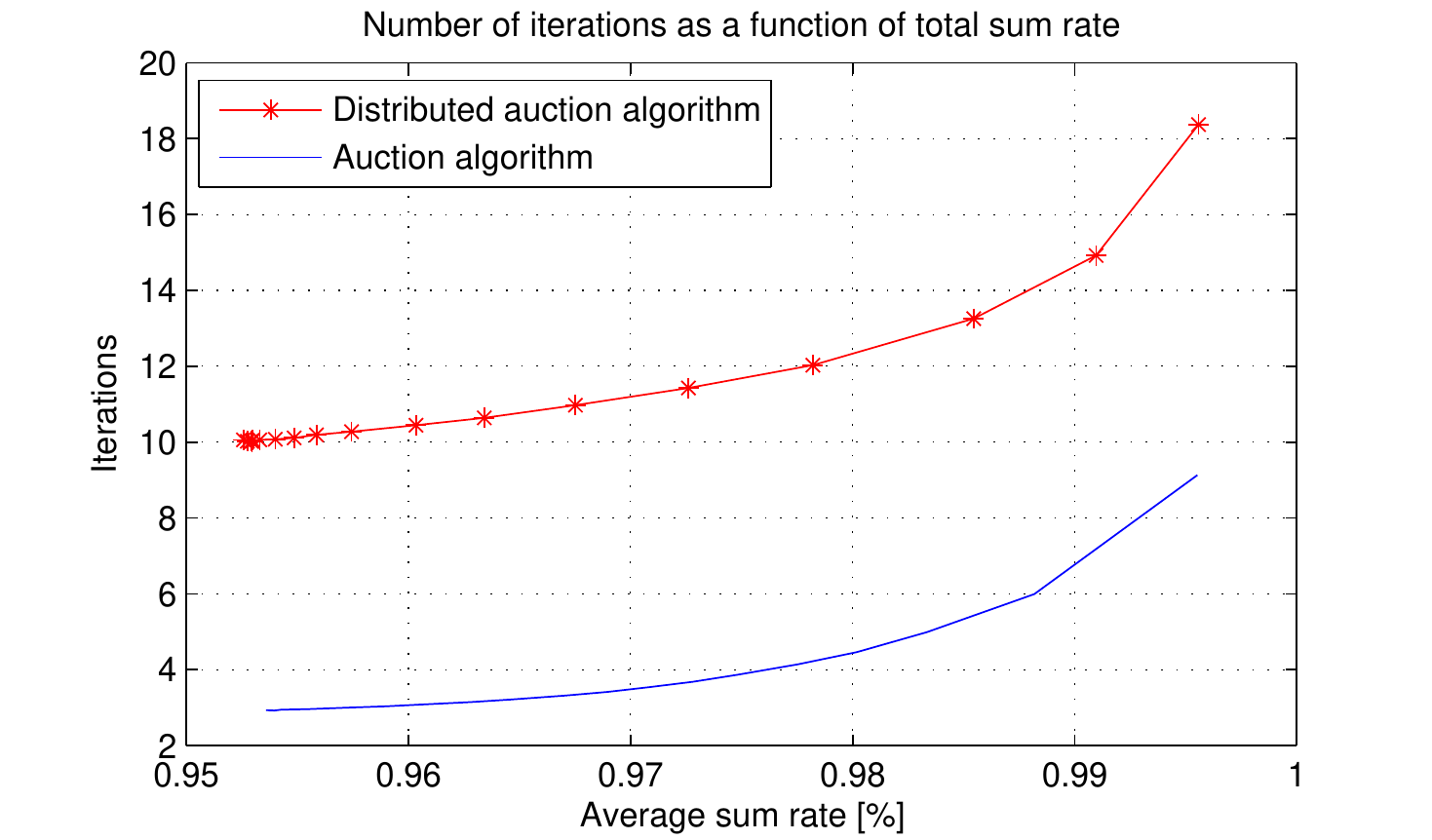}
\caption{Average number of iterations as a function of sum-rate}
\label{figure_iter_comp_opt}
\end{figure}

%\begin{figure}[htbp]
%\centering \epsfig{file=figures/iter_ub2.eps,width=.5\textwidth}
%\caption{Average Number of iterations as a function of $\epsilon$}
%\label{figure_iter_comp}
%\end{figure}
%In Fig. \ref{figure_iter_comp} we compare between the expected convergence times of the distributed auction algorithm for different choices of $\epsilon$ and the upper bound. Although we proved the upper bound only for the case of very small or very large $\epsilon$, the simulations show that the upper bound is valid for arbitrary $\epsilon$.
Fig. \ref{figure_iter_comp_opt} shows the average number of iterations until convergence as a function of the sum-rate achieved by the algorithms.
The average number of iterations needed for the distributed auction algorithm to achieve the same performance as the auction algorithm is only larger by a small constant than the number of iterations needed for the auction algorithm for convergence.
This suggests that the number of iterations needed by both algorithms to converge has the same order of magnitude.
Note that different values of $\epsilon$ should be picked by both algorithms to get the same results.

\begin{figure}[htbp]
\centering \includegraphics[width=0.5\textwidth]{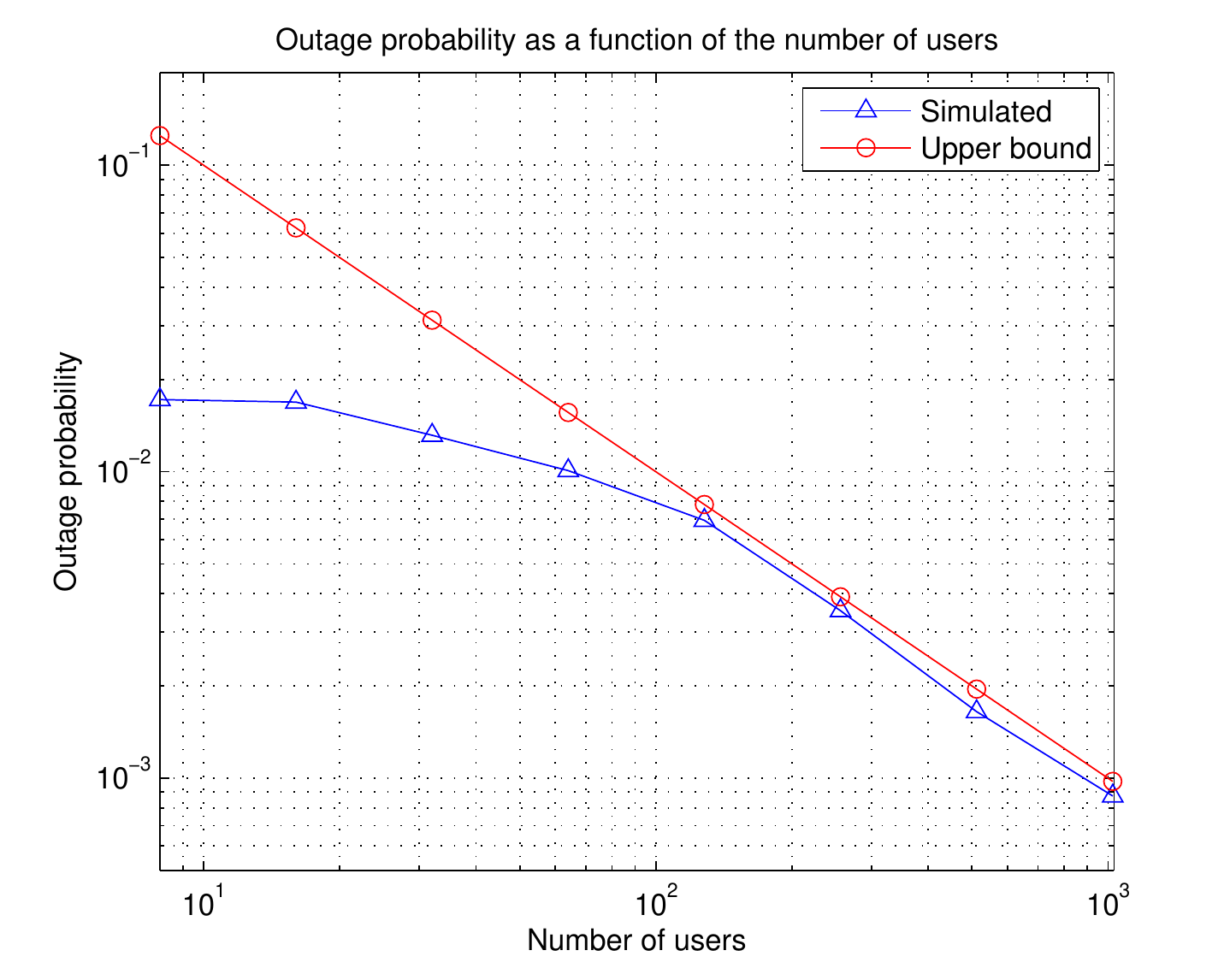}
\caption{Probability that the optimal assignment has at least one entry which is not one of the best $2\log_2(N)$
 order statistics  }
\label{fig_outage}
\end{figure}

\begin{figure}[htbp]
\centering \includegraphics[width=0.5\textwidth]{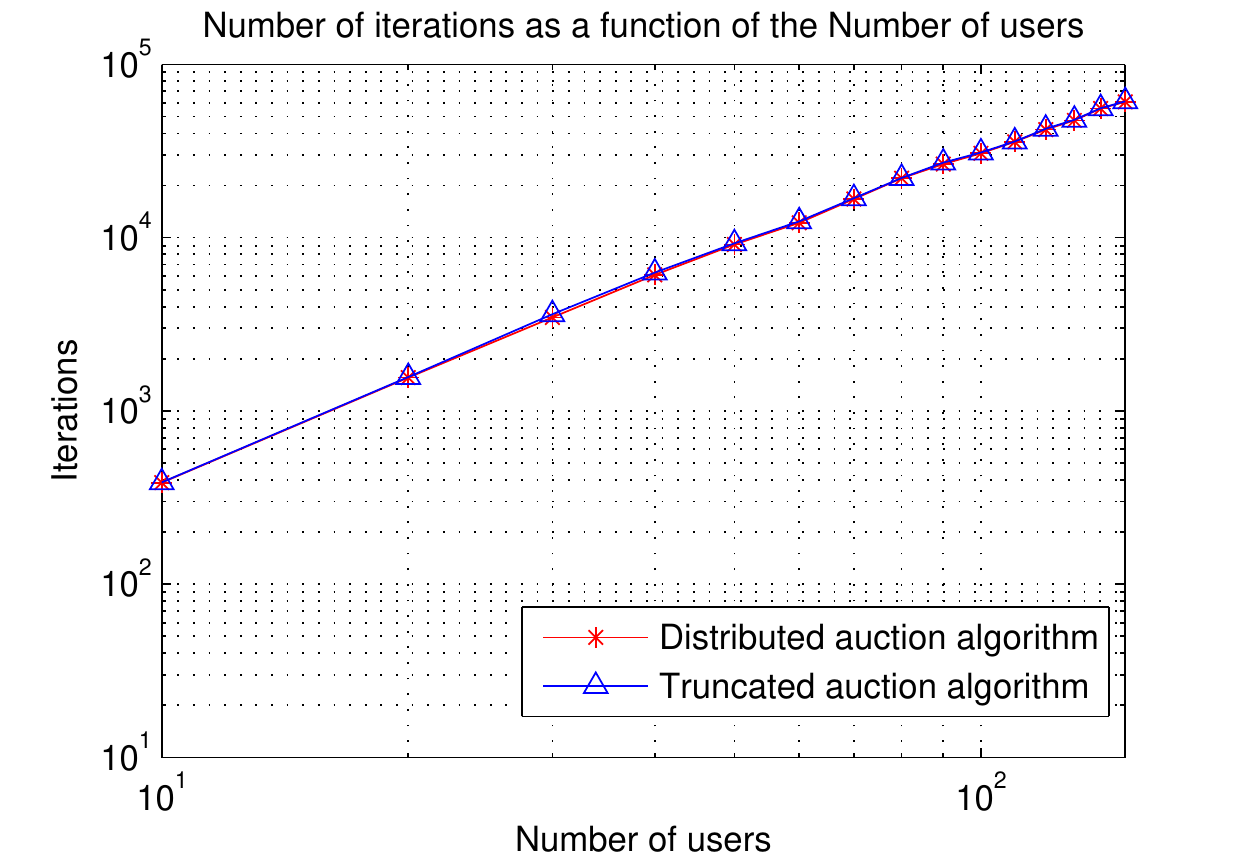}
\caption{Expected number of iterations of the distributed auction algorithm and the truncated auction algorithm.}
\label{fig_variants_iter}
\end{figure}

In the next simulations we compare the performance of the truncated auction algorithm and the distributed auction algorithm.
Fig. \ref{fig_outage} shows the comparison between the outage probability and the upper bound on outage probability as a function of the number of users $N$ using only the best $2\log_2(N)$ order statistics.
As expected the probability is lower than $\frac{1}{N}$.
%Figure. \ref{fig_variants} shows a comparison between the average sum-rate achieved by the distributed auction algorithm and its variants relative to the optimal average sum-rate. The simulations show that the loss tends to zero as the number of users grow.
%This is expected since the variants of the distributed auction algorithm are asymptotically optimal.
%\begin{figure}[htbp]
%\centering \epsfig{file=figures/compare_profit2.eps,width=.5\textwidth}
%\caption{Expected relative reward of the Distributed auction algorithm and its variants.
% order statistics  }
%\label{fig_variants}
%\end{figure}
Figure. \ref{fig_variants_iter} shows the average number of iterations needed for the distributed auction algorithm and the truncated
 auction algorithm until convergence.
The average number of iterations in the truncated auction algorithm and the distributed auction algorithm is almost the same.
This implies that although it was not proven in this paper, upper bounds on the number of iterations in the truncated auction algorithm also hold for the distributed auction algorithm with a full rate matrix.

\begin{figure}[htbp]
\centering \includegraphics[width=0.5\textwidth]{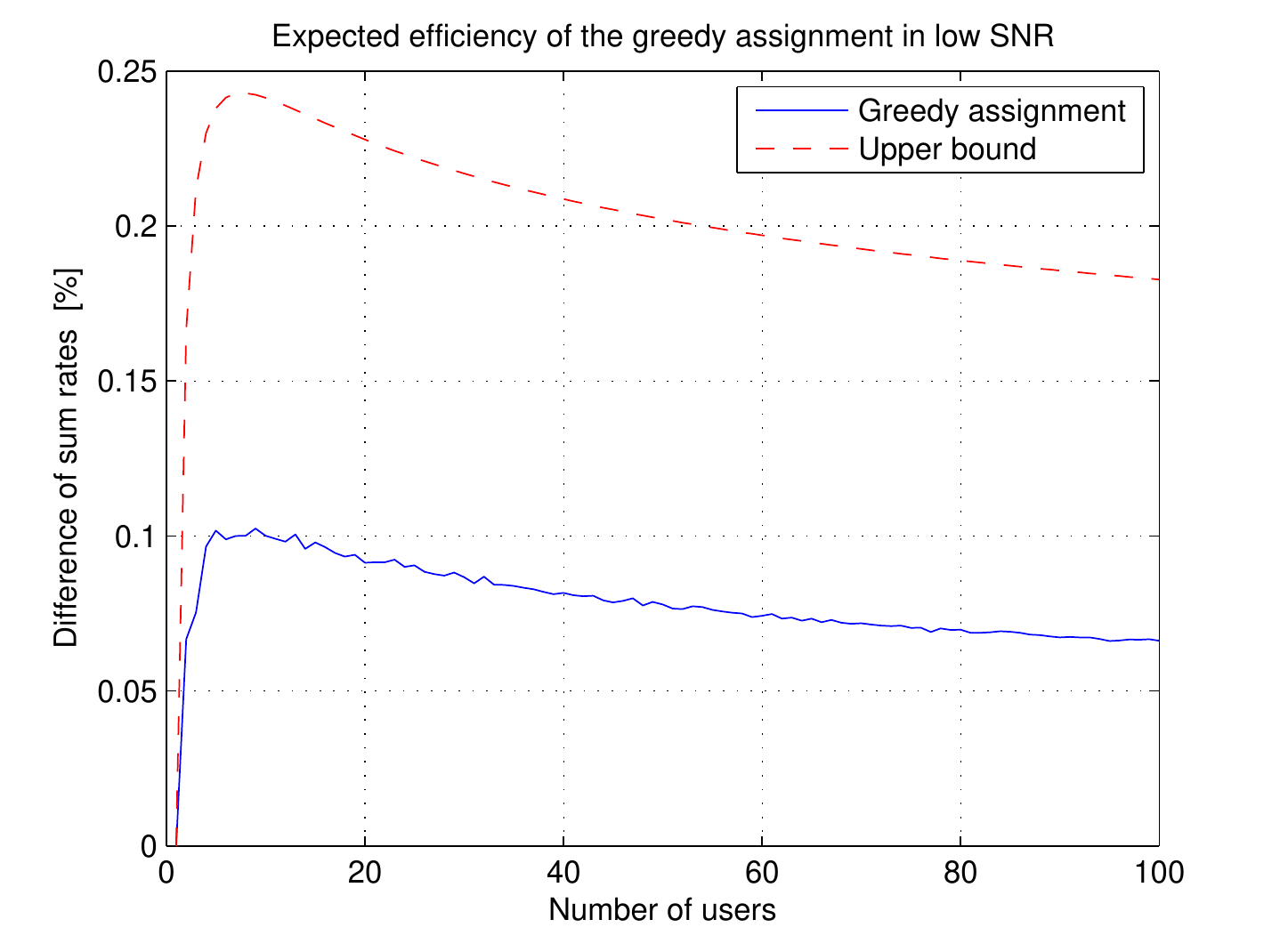}
\caption{Upper bound on the relative error on the expected optimal assignment in low $\SNR$.}
\label{fig_low_snr}
\end{figure}

\begin{figure}[htbp]
\centering \includegraphics[width=0.5\textwidth]{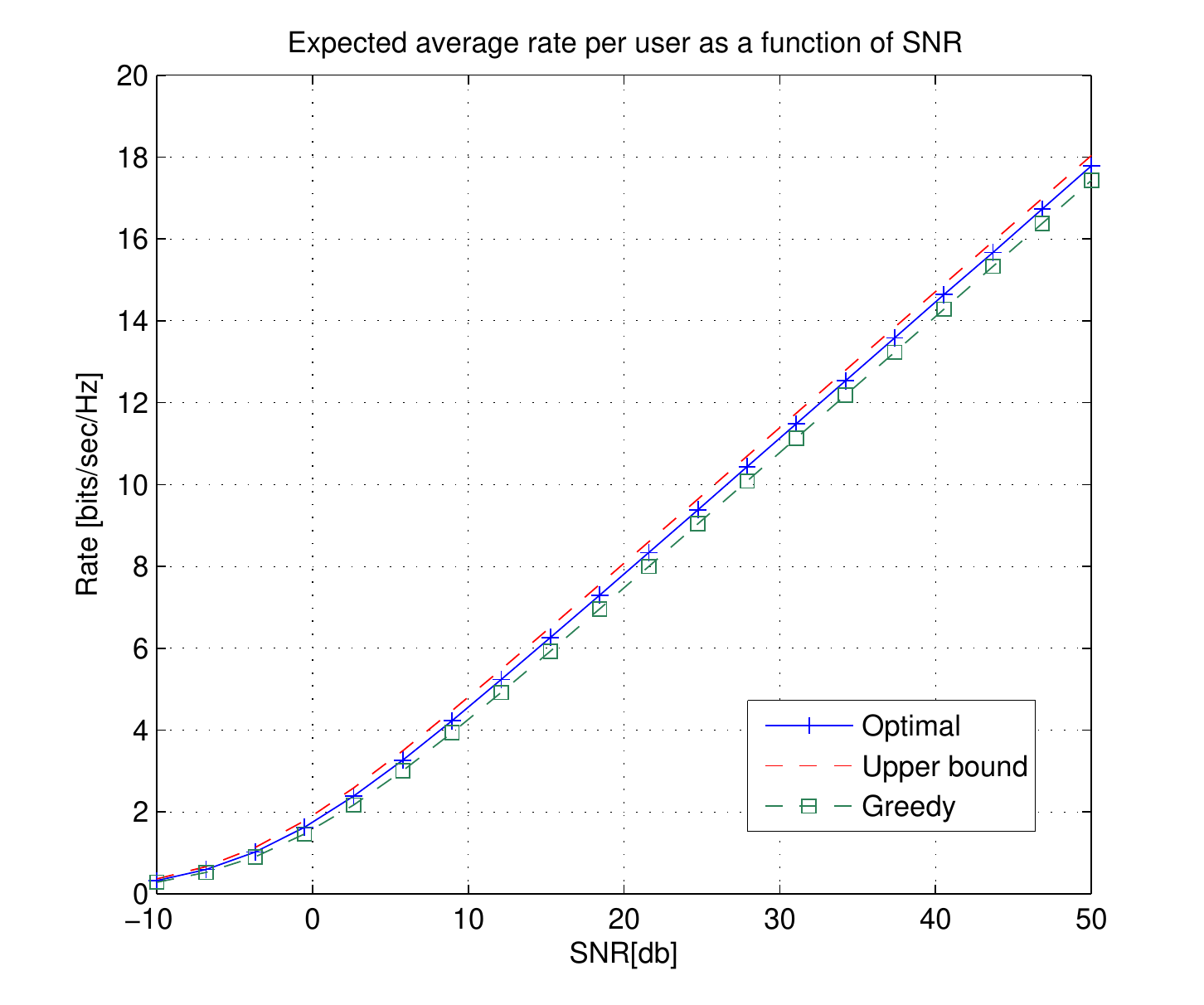}
\caption{Comparison between the average optimal channel assignment and the bounds.}
\label{fig_stable_opt}
\end{figure}

In the next set of simulations we show a comparison between the expected rates achieved by optimal assignment and the rates achieved by the greedy algorithm in i.i.d Rayleigh fading channels.
Fig \ref{fig_low_snr} shows the upper bound on the expected relative error of bounds in the low $\SNR$ scenario. %Note that although the upper bound is asymptotically tight. The real relative loss is much less than the bound.
In the next set of simulations we simulated a system of $10$ channels and $10$ users with i.i.d Rayleigh channels. Fig \ref{fig_stable_opt} compares the optimal channel assignment and the greedy scheme.
As expected, the difference between the upper bound and the greedy bound is constant in high $\SNR$ and converges to $0$ at low $\SNR$.
%Fig \ref{fig_asymptotic} shows the difference between the upper and lower bounds along with the calculated asymptotic constant difference between the bounds. As expected, the difference between the bounds converges to the calculated constant.
%\begin{figure}[htbp]
%\centering \epsfig{file=figures/asymptotic.eps,width=.5\textwidth}
%\caption{Asymptotic behavior of the difference between the upper and lower bounds}
%\label{fig_asymptotic}
%\end{figure}

\begin{figure}[htbp]
\centering \includegraphics[width=0.5\textwidth]{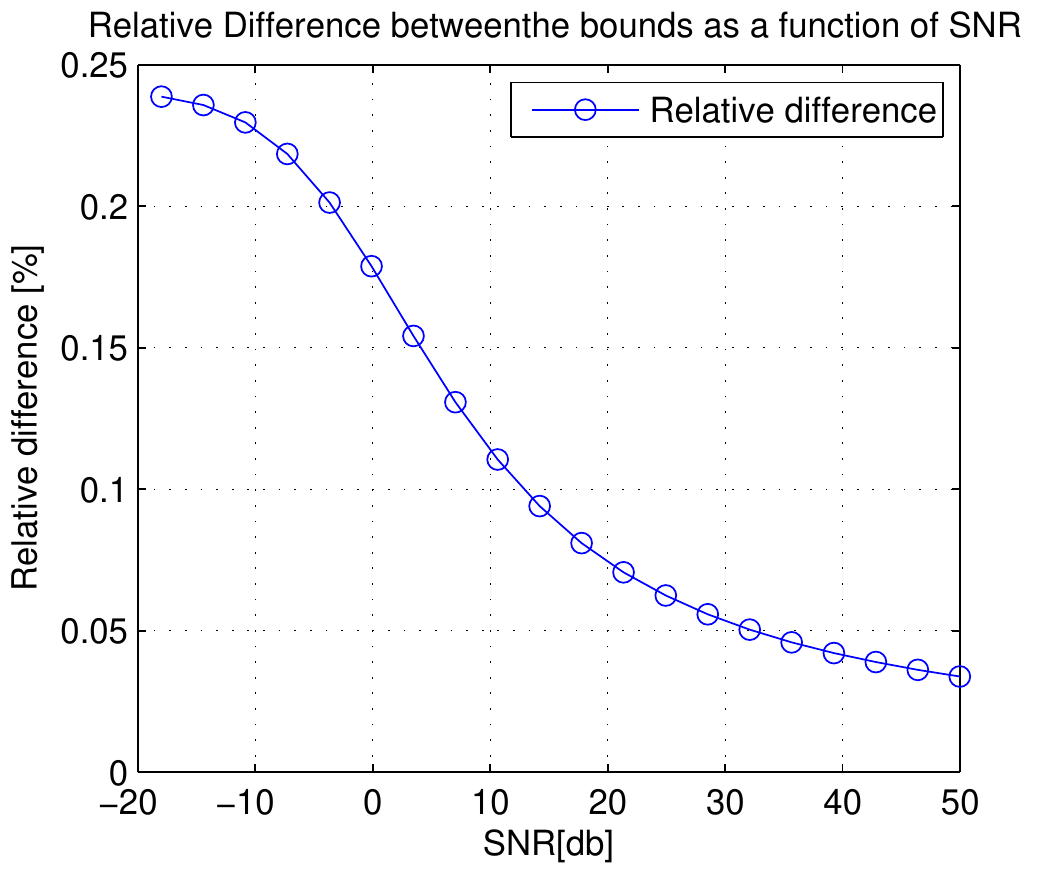}
\caption{Asymptotic behavior of the relative difference between the upper and lower bounds}
\label{fig_asymptotic_low}
\end{figure}

Fig \ref{fig_asymptotic_low} show the relative difference between the bounds and the asymptotic relative difference in low $\SNR$.
One can easily see that as expected, the relative difference goes to zero as $\SNR$ increases. Furthermore, at a $\SNR$ of $30db$, which is the average working $\SNR$ of WiFi networks, the average sum-rate of the greedy scheme is $95\%$ of the upper bound on the optimal assignment.

\section{Conclusion}
\label{section_conclusions}
In this paper the problem of fully distributed channel assignment was addressed.
The distributed auction algorithm was introduced. The distributed auction algorithm can be implemented in a fully distributed manner and is  $\epsilon$-optimal.
We showed that the distributed auction algorithm shares the same optimality properties as the original auction algorithm.
A variant of the distributed auction algorithm dubbed the truncated auction algorithm was introduced and was shown to be asymptotically optimal under mild conditions on the distribution of the random matrix.
 We then analyzed the optimal assignment in i.i.d Rayleigh fading channels and showed that a randomized greedy scheme is asymptotically optimal with a large number of users or when the $\SNR$ is sufficiently high.
We then showed simulated results of the algorithms developed in here and compared their performance with the upper and lower bounds derived in the paper.

\appendix
\subsection{Proof of theorem \ref{theorem_finite_time1}}
\label{appendix_proof1}
In this appendix we follow the proof from \cite{bertsekas1992auction} and show that for a feasible problem and for any positive value of $\epsilon$, the distributed auction algorithm terminates with a feasible assignment which is within $M\epsilon$ of being optimal in a finite number of iterations.
We assume in this proof that the number of people is at most equal to the number of objects, as this condition can always be achieved by adding dummy zero objects.
The proof relies on the following facts:
\begin{enumerate}
\item In the auction algorithm as well in the distributed auction algorithm a bidding and assignment
phase can result in a reassignment of an already assigned object to a different person, but cannot
result in the object becoming unassigned. Therefore when an object is assigned, it remains assigned throughout the remainder of the algorithm's duration. Furthermore, except at termination, there will always exist at least one object that has never been
assigned, and has a price equal to its initial price.
\item Each time an object receives a bid, its local price increases by at least $\epsilon$. Therefore, if an
object receives a bid an infinite number of times, its price increases to $\infty$ for at least one person.
\item Every $|A(i)|$ bids per person $m$, where $A(i)$ is the set of objects user $i$ can make bids on and $|A(i)|$ is the number of objects in $A(i)$; the best object
value $\nu_i$ defined by
\beq
\nu_i=\max_{j\in A(i)}(a_{ij} - p_{ij})
\eeq
decreases by at least $\epsilon$. The reason is that a bid by person $i$ either decreases $\nu_i$ by at least $\epsilon$, or else leaves $\nu_i$ unchanged because there is more than one object $j$ attaining the maximum.
However, in the latter case, the price of the object $j_i$ receiving the bid will increase by at least $\epsilon$,
and object $j_i$ will not receive another bid by person $i$ until $\nu_i$ decreases by at least $\epsilon$. The conclusion
is that if a person $i$ bids an infinite number of times, $\nu_i$ must decrease to $-\infty$.
\end{enumerate}
We now argue by contradiction. If termination did not occur, the subset $J^{\infty}$ of objects that received
an infinite number of bids is nonempty. Also, the subset of persons $I^{\infty}$ that bid an infinite number of
times is nonempty. As argued in 2), the prices of the objects in $J^{\infty}$ must tend to $\infty$, while as
argued in 3) above, the scalars $\nu_i=\max_{j\in A(i)}(a_ij - p_ij)$ must decrease to $-\infty$ for all persons $i\in I^{\infty}$. Therefore, $a_{ij}- p_{ij}$ tends to $-\infty$ for all $j \in A(i)$, implying that
\beq
A(i)\subset J^{\infty}, \forall i\in I^{\infty}
\eeq
The L$\epsilon$-CS condition states that $ a_{ij}- p_{ij} \geq\nu_i+\epsilon$ for every assigned pair (i, j); thus after a finite number
of iterations, each object in $J^{\infty}$ can only be assigned to a person from $I^{\infty}$. Since after a finite number
of iterations at least one person from $I^{\infty}$ will be unassigned at the start of each iteration, it follows that
the number of people in $I^{\infty}$ is strictly larger than the number of objects in $J^{\infty}$. This contradicts the
existence of a feasible assignment, since people in $I^{\infty}$ can only be assigned to objects in $J^{\infty}$.
Therefore, the algorithm must terminate. The feasible assignment obtained upon termination satisfies
L$\epsilon$-CS (since the algorithm preserves L$\epsilon$-CS throughout); hence by theorem \ref{theorem1}, this assignment is within $M\epsilon$ of being optimal.

\bibliographystyle{ieeetr}

\begin{biography}[{\includegraphics[width=1in,height =1.25in,clip,keepaspectratio]{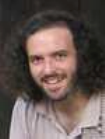}}]{Oshri Naparstek}
received his B.Sc. in applied mathematics in 2006 and his M.Sc. (cum laude) in applied mathematics in 2009 from Bar-Ilan University, Ramat-Gan, Israel.
He is currently working towards his Ph.D. degree in the Faculty of Engineering, Bar-Ilan University.

His main research interests include distributed optimization techniques, communication protocols, and dynamic spectrum management for wireless and wireline networks.
\end{biography}
\begin{biography}[{\includegraphics[width=1in,height =1.25in,clip,keepaspectratio]{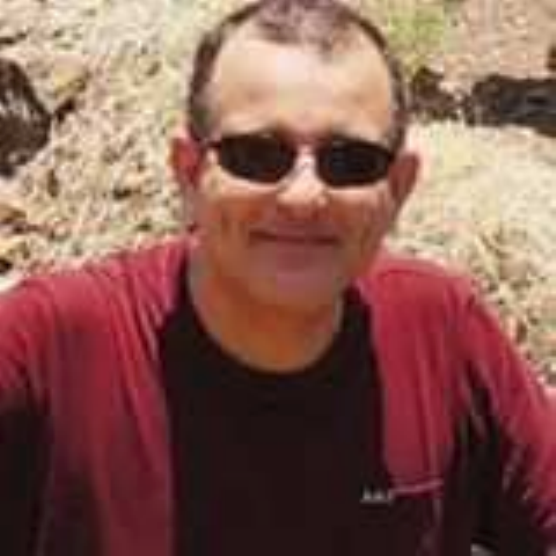}}]{Amir Leshem}
(M’98–SM’06) received the B.Sc.
(cum laude) in mathematics and physics, the
M.Sc. (cum laude) in mathematics, and the Ph.D.
degree in mathematics all from the Hebrew University, Jerusalem, Israel, in 1986, 1990, and 1998,
respectively.
From 1998 to 2000, he was with the Faculty of
Information Technology and Systems, Delft University of Technology, The Netherlands, as a postdoctoral fellow working on algorithms for the reduction
of terrestrial electromagnetic interference in radio-astronomical radio-telescope antenna arrays and signal processing for communication. From 2000 to 2003, he was Director of Advanced Technologies with
Metalink Broadband where he was responsible for research and development
of new DSL and wireless MIMO modem technologies and served as a member
of ITU-T SG15, ETSI TM06, NIPP-NAI, IEEE 802.3 and 802.11. From 2000
to 2002, he was also a visiting researcher at Delft University of Technology. He
is a Professor and one of the founders of the faculty of engineering at Bar-Ilan
University where heads the Signal Processing track. From 2003 to 2005, he was
the technical manager of the U-BROAD consortium developing technologies to
provide 100 Mbps and beyond over copper lines. His main research interests include multichannel wireless and wireline communication, applications of game
theory to dynamic and adaptive spectrum management of communication networks, array and statistical signal processing with applications to multiple element sensor arrays and networks, wireless communications, radio-astronomical
imaging and brain research, set theory, logic and foundations of mathematics.
Dr. Leshem was an Associate Editor of the IEEE TRANSACTIONS ON SIGNAL
PROCESSING from 2008 to 2011, and he was the Leading Guest Editor for
special issues on signal processing for astronomy and cosmology in IEEE
SIGNAL PROCESSING MAGAZINE and the IEEE JOURNAL OF SELECTED TOPICS
\end{biography}

\end{document}